\documentclass[a4paper,12pt]{article}
\pdfoutput=1
\usepackage{amsfonts,amsmath,amssymb}
\usepackage{graphicx}
\usepackage{color,here}
\usepackage{cite}
\usepackage{hyperref}
\usepackage{finite_GGB}
\usepackage{simplewick}
\usepackage{enumitem}
\usepackage{mathrsfs,mathdots,stmaryrd}
\usepackage[vcentermath]{youngtab}
\usepackage{tocloft}

\usepackage{pgf}
\usepackage{ytableau,varwidth}

\usepackage{xparse}
\usepackage{tikz}
\usetikzlibrary{arrows,arrows.meta,backgrounds,calc,cd,decorations.markings,decorations.pathmorphing,decorations.shapes,external,graphs,math,matrix,positioning,shapes.geometric}

\numberwithin{equation}{section}

\let\OLDthebibliography\thebibliography
\renewcommand\thebibliography[1]{
  \OLDthebibliography{#1}
  \setlength{\parskip}{2pt}
  \setlength{\itemsep}{2pt plus 2pt}
}

\begin{document}
\thispagestyle{empty}

\ \vskip 30mm

{\LARGE 
\centerline{\bf Oscillating Multiple Giants}

}

\vskip 25mm

\centerline{
{\large \bf Ryo Suzuki}
}

{\let\thefootnote\relax\footnotetext{{\tt rsuzuki.mp\_at\_gmail.com}}}

\vskip 15mm

\centerline{{\it Shing-Tung Yau Center of Southeast University,}}
\centerline{{\it 15th Floor, Yifu Architecture Building, No.2 Sipailou,}}
\centerline{{\it Xuanwu district, Nanjing, Jiangsu, 210096, China}}

\vskip 25mm


\centerline{\bf Abstract}

\vskip 10mm 

We propose a new example of the AdS/CFT correspondence between the system of multiple giant gravitons in \AdSxS\ and the operators with $O(N_c)$ dimensions in $\cN=4$ super Yang-Mills. 
We first extend the mixing of huge operators on the Gauss graph basis in the $\alg{su}(2)$ sector to all loops of the 't Hooft coupling, by demanding the commutation of perturbative Hamiltonians in an effective $U(p)$ theory, where $p$ corresponds to the number of giant gravitons.
The all-loop dispersion relation remains gapless at any $\lambda$, which suggests that harmonic oscillators of the effective $U(p)$ theory should correspond to the classical motion of the D3-brane that is continuously connected to non-maximal giant gravitons.

\newpage
\tableofcontents

\section{Introduction}\label{sec:intro}

Traditionally the AdS/CFT correspondence has been studied in the planar large $N_c$ limit \cite{Maldacena:1997re}.
Whether AdS/CFT holds true in a non-planar but still large $N_c$ limit is a challenging question.
Such a nontrivial limit can be implemented by deforming the background theory or spacetime, or by introducing semiclassical objects carrying the dimensions or energy of order $N_c$\,.

One of the most studied examples of AdS/CFT is the correspondence between string theory on \AdSxS\ and $\cN=4$ super Yang-Mills (SYM) theory. We depart from the planar region of $\cN=4$ SYM by studying huge operators whose dimensions are comparable to the rank of the gauge group $N_c$\,.\footnote{Another promising approach to a non-planar large $N_c$ limit is the localization, which is valid at any $N_c$ and can extract non-BPS data \cite{Binder:2019jwn}.}
The operators with $O(N_c^2)$ dimensions correspond to the Lin-Lunin-Maldacena (LLM) geometry at strong coupling \cite{Lin:2004nb}, while those with $O(N_c)$ dimensions correspond to the giant graviton \cite{McGreevy:2000cw}.
This correspondence continues to non-BPS operators in both of the $O(N_c^2)$ and $O(N_c)$ cases.
For the former case, an isomorphism between non-BPS states was conjectured between the LLM geometry and $\cN=4 $ SYM \cite{Koch:2016jnm,deMelloKoch:2018tlb,deMelloKoch:2018ert,Berenstein:2020jen,Suzuki:2020oce}.
For the latter case, non-BPS states around the giant graviton are less well-understood.
This is the main subject of this paper.

\bigskip
Let us first review recent progress on the weak coupling side.

In AdS/CFT, the half-BPS operators with huge dimensions should be organized through the operator basis labeled by a Young diagram \cite{Corley:2001zk}. Similarly, a convenient way to describe non-BPS operators with huge dimensions is the restricted Schur basis, labeled by a set of Young diagrams \cite{Balasubramanian:2004nb,deMelloKoch:2007rqf}. 
The dilatation operator expressed in this basis mixes the Young diagrams with different shapes.

For simplicity, consider the operators in the $\alg{su}(2)$ sector, which consists of complex scalars $Z$ and $Y$ of $\cN=4$ SYM. Suppose that a small number of $Y$'s are added to a large number of $Z$'s. 
If the Young diagram representing $Z$'s has $p$ long columns, this type of operators roughly corresponds to a system of $p$ spherical giant gravitons in \AdSxS. The Young diagram representing $Y$'s describes a small fluctuation of the giant gravitons.\footnote{Here $Y$'s and $Z$'s constitute a huge Young diagram whose shape wildly fluctuates due to the operator mixing. This situation is different from a single-trace operator coupled to $\det (Z)$, where the operator mixing does not spoil the color structure at the leading order of large $N_c$\,.}

The one-loop mixing in this setup is remarkably simple.
First, the number of columns $p$ does not change at large $N_c$\,, because giant gravitons are semi-classical objects at strong coupling \cite{DeComarmond:2010ie}. Second, the operator mixing splits into the mixing of $Z$'s and the mixing of $Y$'s. Third, the one-loop spectrum eventually reduces to a set of $p$ decoupled harmonic oscillators \cite{Carlson:2011hy,Koch:2011hb}. The last observation is called the non-planar integrability in the literature.

We should emphasize that the mixing problem of huge operators is quite different from the planar mixing problem, and the development of sophisticated techniques has been crucial.
The mixing problem of $Y$'s is solved by the Gauss graph basis, which counts the number of open strings ending on different giant graviton branes \cite{deMelloKoch:2012ck}. 
The same technique can be used to simplify the mixing in the $\alg{su}(3)$ and $\alg{su}(2|3)$ sectors \cite{Koch:2011jk,Koch:2012sf}.
Explicit computation of the mixing matrix has been given up to two loops in $g_{\rm YM}^2$ in \cite{deMelloKoch:2012sv}. There are also trial studies at higher loops \cite{Koch:2013xaa,Lin:2014yaa}.
More generally, the mixing problem corresponding to $p$ giant gravitons can be described by an effective $U(p)$ theory \cite{deCarvalho:2018xwx,deMelloKoch:2020agz}.
The Hamiltonian of the effective $U(p)$ theory has the symmetry $U(1)^p$ in the distant corners approximation, namely when the differences of the length of the adjacent columns are large.

\bigskip
Next, the key developments on the strong coupling side are summarized.

The D-brane motion is described by a low energy effective action which consists of Dirac-Born-Infeld (DBI) and Chern-Simons (CS) terms \cite{Polchinski:1996na}.
The giant graviton is a classical solution of the D3-brane action moving in the \AdSxS\ background. The spherical giant graviton wraps ${\rm S}^3$ inside ${\rm S}^5$ \cite{McGreevy:2000cw}, and the AdS giant wraps  ${\rm S}^3$ inside ${\rm AdS}_5$ \cite{Grisaru:2000zn,Hashimoto:2000zp}. 
The quantum fluctuation modes around the giant graviton have been studied in \cite{Das:2000st}.

The open strings ending on the giant graviton have been studied from two viewpoints.
In the first viewpoint, we replace open strings with $U(1)$ flux and study the D-brane. The classical motion of a D3-brane in such a background spacetime has been studied in the flat space \cite{Callan:1997kz}, in the pp-wave \cite{Sadri:2003mx} and in \AdSxS\ \cite{Hirano:2006ti}. The $U(1)$ gauge fields typically become spiky, and they diverge at the location where open strings end on the D-brane.
In the second viewpoint, we study the open string as a classical integrable system \cite{Dekel:2011ja}, or a boundary integrable system \cite{Hofman:2007xp}.\footnote{A coherent state description of open strings ending on a mixture of (generally non-maximal) giant gravitons is studied in \cite{Berenstein:2013md,Berenstein:2013eya,Berenstein:2014zxa}.}
In both points of view, the brane-string system typically has divergent energy, which is canceled by a divergent angular momentum of open strings, just like the giant magnon \cite{Hofman:2006xt}.

\bigskip
It is expected that the system of open strings with $p$ giant gravitons corresponds to the effective $U(p)$ theory, but there is still obscurity in this understanding as AdS/CFT. 
The purpose of this paper is to understand this theory more precisely by revisiting the analysis both in gauge and string theories.

In Section \ref{sec:mixing N=4} and \ref{sec:all-loop}, we study the perturbative Hamiltonians of the effective $U(p)$ theory on the gauge theory side in detail. 
Possible forms of the effective Hamiltonian are constrained by the $GL(p)$ algebra, and by demanding that the perturbative Hamiltonians at each loop order commute with each other.
We find that there are at most $(\ell+1)$ linearly-independent mutually-commuting operators at $\ell$-loops.
In the continuum limit, these candidate operators reduce to the harmonic oscillators at one loop, which allows us to conjecture an all-loop ansatz,
\begin{equation}
\Delta - J = \frac{\tilde f (\lambda)}{N_c} \, m \, n_{12} (\sigma), \qquad
n_{12} (\sigma) \in \bb{Z}_{\ge 0} \,, \qquad
m = 1,2, \dots, \left\lceil N_c - \frac{n_Z}{2} + 1 \right\rceil 
\label{intro all-loop}
\end{equation}
where we put $p=2$ for simplicity, and $\tilde f (\lambda)$ is an unknown function of the 't Hooft coupling $\lambda = N_c \, g_{\rm YM}^2$\,.

Our ansatz \eqref{intro all-loop} predicts two remarkable consequences. First, the anomalous dimensions remain non-zero at the leading order of large $N_c$\,, because $m_\alpha$ can be an integer of $O(N_c)$.
Second, the excitations are gapless.
Recall that the energy of an open string attached on the $Z=0$ giant gravitons is gapped, because the open string stretching on S$^5$ carries non-zero energy, equal to the length times tension.\footnote{Here the energy is measured in the unit of string tension $\sqrt{\lambda}$ and not in the D3-brane tension $N_c/g_s \sim N_c^2/\lambda$. The string with a finite length produces a gap in the dispersion relation, even if $g_s \ll 1$ or $\lambda \ll N_c$\,.}
This disagreement indicates that the previous analyses on the LLM geometry do not immediately apply to the system of multiple giant gravitons. 
What does the all-loop ansatz represent at strong coupling?

In Section \ref{sec:strong coupling} we revisit a classical single D3-brane wrapping S$^3$ inside \AdSxS, and solve the equations of motion around the BPS spherical giant graviton.
Following the steps similar to the stability analysis of \cite{Das:2000st}, we found two types of classical solutions oscillating around the BPS giant gravitons.
The first type is a point-like D-brane, and the second type is a fuzzy D-brane with non-trivial KK modes on S$^3$. The energy of the latter solution is
\begin{equation}
E - J = \frac{N_c}{g_s} \[
\epsilon^2 \, \frac{c_k^2 \, (k+1)^2}{8 (1-j) (k+2)} 
+ O(\epsilon^4) \]
\label{intro KK energy}
\end{equation}
where $c_k$ is a numerical constant that remains finite as $k \to \infty$.
We argue that the latter solution \eqref{intro KK energy} is a good candidate for the string theory state corresponding to the all-loop harmonic oscillator \eqref{intro all-loop} at strong coupling.
Our reasoning will be presented in Section \ref{sec:AdS/CFT}.

\bigskip
This paper is supplemented by {\tt Mathematica} files used for the computations in Sections \ref{sec:all-loop}, \ref{sec:strong coupling}.

\section{Mixing of huge operators in $\cN=4$ SYM}\label{sec:mixing N=4}

We collect known facts about the perturbative mixing of huge operators in $\cN=4$ SYM.
Our notation and basis facts about the Gauss graph basis are summarized in Appendix \ref{app:Gauss}.

\subsection{Perturbative dilatation operator}\label{sec:pert dil}

We express the perturbative dilatation operator in the $\alg{su}(2)$ sector of $\cN=4$ SYM by
\begin{equation}
D (g_{\rm YM}) =\sum_{\ell=0}^{\infty} \( \frac{g_{\rm YM}}{4 \pi} \)^{2 \ell} D_\ell
\label{def:DgYM}
\end{equation}
where \cite{Beisert:2003tq}
\begin{align}
D_0 &= \Tr Y\check Y + \Tr Z\check Z 
\label{BKS D0} \\[1mm]
D_1 &=
-2 :\! \Tr [Y, Z][\check Y, \check Z] \!:\! 
\label{BKS D1} \\[1mm]
D_2 &=
-2 :\! \Tr \Big[ [Y, Z] ,\check Z  \Big] \Big[ [\check Y, \check Z] , Z  \Big] \!:\!
- 2 :\! \Tr \Big[ [Y, Z], \check Y  \Big] \Big[ [\check Y, \check Z], Y  \Big] \!:\!\
- 2 \(N_c-1\) D_1 \,.
\label{BKS D2}
\end{align}
The fields $(\Phi , \check \Phi)$ satisfy the $U(N_c)$ Wick rules
\begin{equation}
\Tr (A \check \Phi B \Phi ) = \Tr (A) \, \Tr(B), \qquad
\Tr (A \check \Phi) \, \Tr ( B \Phi ) = \Tr (A B) , \qquad
\Tr (1) = N_c \,.
\end{equation}
The explicit form of $D_\ell$ has been known up to five loops \cite{Beisert:2007hz,Fiamberti:2009jw},

We assume $n_Z = O(N_c), \ n_Y = O(1)$. The operators in the Gauss graph basis are denoted by
\begin{equation}
O^{R,r}( \sigma) = O (\vec l) = O(l_1, l_2 , \dots, l_p)
\end{equation}
as in \eqref{app:Ovec l}, where $\vec l$ specifies the Young diagram for $Z$.
Let us write $D_\ell$ acting on the Gauss graph operators by $D_\ell^G$\,.
At the leading order of large $N_c$\,, these dilatation operators are given by \cite{DeComarmond:2010ie,deMelloKoch:2011ci,deMelloKoch:2012ck,deMelloKoch:2012sv}
\begin{equation}
D_1^G = - \sum_{\substack{i,j=1 \\ i \neq j}}^p n_{ij} ( \sigma ) \Delta_{ij}^{(1)} \, \qquad
D_2^G = - \sum_{\substack{i,j=1 \\ i \neq j}}^p n_{ij} ( \sigma ) \pare{ (L - 2 N_c) \, \Delta_{ij}^{(1)}+\Delta_{ij}^{(2)} }
\label{Dil12 GGB}
\end{equation}
where\footnote{We slightly modified the result of \cite{deMelloKoch:2012sv} which computed only the first term of $D_2$ in \eqref{BKS D2}. We added the third term of $D_2$ and symmetrized $\Delta^{(2)}_{ij}$ with respect to $i \leftrightarrow j$, owing to $n_{ij} ( \sigma ) = n_{ji} ( \sigma )$. Note that we sum over $i \neq j$ and not $i < j$.}
\begin{equation}
\begin{aligned}
\Delta^{(1)}_{ij} &= \Delta^+_{ij} +\Delta^0_{ij} + \Delta^-_{ij}
\\[1mm]
\Delta^{(2)}_{ij} &= (\Delta^+_{ij})^2 +\Delta^0_{ij}\Delta^+_{ij} + \Delta^+_{ij}\Delta^-_{ij}
+ \Delta^-_{ij}\Delta^+_{ij} +\Delta^0_{ij}\Delta^-_{ij} +(\Delta^-_{ij})^2 \,.
\end{aligned}
\label{two-loop Delta ij}
\end{equation}
The difference operators $\Delta_{ij}^0, \Delta_{ij}^\mp$ are defined by
\begin{equation}
\begin{aligned}
\Delta_{ij}^0 \, O (\vec l) &= - \Big( h(i, l_i) + h(j, l_j) \Big) \, O (\vec l)
\\[1mm]
\Delta_{ij}^- \, O (\vec l) &= \sqrt{ h(i, l_i) \, h(j, l_j+1)} \, O( \dots, l_i-1, \dots ,l_j+1, \dots)
\\[1mm]
\Delta_{ij}^+ \, O (\vec l) &= \sqrt{ h(i, l_i+1) \, h(j, l_j)} \, O( \dots, l_i+1, \dots ,l_j-1, \dots) 
\end{aligned}
\label{def:Delta ij}
\end{equation}
where $h(i, l_i)$ is the box weight for spherical giants,
\begin{equation}
h(i, l_i) \equiv N_c + i - l_i \,, \qquad
h(i, l_i \mp 1) - h(i, l_i) = \pm 1.
\label{def:hili}
\end{equation}
We find it convenient to keep $O(1)$ terms in \eqref{def:Delta ij} although $h(i, l_i) \sim h(i, l_i \mp 1)$ at large $N_c$\,.
An important feature of $D_\ell^G$ is that it consists of a sum over a pair of indices $(i,j)$, and the third row/column does not show up.

\bigskip
We can simplify the difference operators in \eqref{def:Delta ij} by introducing 
\begin{equation}
\begin{aligned}
d_i^- \, O (\vec l) &= \sqrt{ h(i, l_i) } \, O( \dots, l_i-1, \dots) 
\\[1mm]
d_i^+ \, O (\vec l) &= \sqrt{ h(i, l_i+1) } \, O( \dots, l_i+1, \dots) 
\\[1mm]
\hat h_i \, O (\vec l) &= h (i, l_i) \, O (\vec l) .
\end{aligned}
\label{def:dpmh} 
\end{equation}
These operators satisfy the relations
\begin{equation}
d_i^+ d_i^- = \hat h_i \,, \qquad
\bigl[ d_i^+ , d_j^- \bigr] = \delta_{ij} 
\label{CCR for d}
\end{equation}
which follow from
\begin{equation}
\[ d_i^+ , d_i^- \] O (\vec l) = \Bigl\{ h(i, l_i) - h(i, l_i+1) \Bigr\} \, O (\vec l)
= O (\vec l).
\end{equation}
Note that
\begin{equation}
\sum_{i=1}^p \hat h(i, l_i) \, O (\vec l) = n_Z \, O (\vec l).
\end{equation}

We can rewrite the difference operators in \eqref{two-loop Delta ij} as
\begin{align}
\Delta^{(1)}_{ij} &= - \( d_i^+ - d_j^+ \) \( d_i^- - d_j^- \) 
\label{Del1 GGB} \\[1mm]
\Delta^{(2)}_{ij} &= - \( d_i^+ - d_j^+ \) \Bigl( 1 + d_i^+ d_j^- + d_j^+ d_i^- \Bigr) \( d_i^- - d_j^- \) .
\label{Del2 GGB}
\end{align}
If we introduce
\begin{align}
\cH_1 &= \sum_{i \neq j}^p n_{ij} (\sigma) \, \cH_{1,ij}
\equiv \sum_{i \neq j}^p n_{ij} (\sigma) \( d_i^+ - d_j^+ \) \( d_i^- - d_j^- \)
\label{def:cH1} \\[1mm]
\cH_2 &= \sum_{i \neq j}^p n_{ij} (\sigma) \, \cH_{2,ij}
\equiv \sum_{i \neq j}^p n_{ij} (\sigma)
\( d_i^+ - d_j^+ \) \Bigl(  d_i^+ d_j^- + d_j^+ d_i^- \Bigr) \( d_i^- - d_j^- \)
\label{def:cH2}
\end{align}
we obtain
\begin{equation}
D_1^G = - \cH_1 \,, \qquad
D_2^G = - \( L - 2 N_c \) \cH_1 - \cH_2 \,.
\label{Del GGB-cH}
\end{equation}
The dilatation operator $D_\ell^G$ written in terms of $\{ d_i^+ , d_i^- \}$ can be regarded as the Hamiltonian of an effective $U(p)$ theory. This is because $E_{ij} \equiv d_i^+ d_j^-$ satisfies the $GL(p)$ commutation relations,\footnote{Roughly speaking, $E_{ij}$ is a $p \times p$ matrix whose entries are zero except at the $i$-th row, $j$-th column. The Hermitian combinations generate $U(p)$.}
\begin{equation}
[ E_{ij} \,, E_{kl} ] = \delta_{jk} \, E_{il} - \delta_{il} \, E_{kj} \,. 
\label{GL(p) commutation}
\end{equation}
We assume that $n_{ij} (\sigma)$ are general non-negative integers. Then the perturbative Hamiltonians in \eqref{def:cH1}, \eqref{def:cH2} are invariant only under $U(1)^p$.

\subsection{Commutation relations}\label{sec:commutation}

The one- and two-loop dilatations at $p=2$ in \eqref{def:cH1}, \eqref{def:cH2} satisfy the relation
\begin{equation}
\[ \cH_{1,ij} \,, \cH_{2,ij}\] = 0
\label{cH12 commute p=2}
\end{equation}
and  the perturbative dilatation operators on the Gauss graph basis commute.
This is a surprising feature because the planar dilatation operators (on the single-trace operators) at one- and two-loop do not commute \cite{Beisert:2003tq}.\footnote{In general, the operator $C(g)=A+gB$ has the eigenvectors independent of $g$ when $[A,B]=0$. Conversely, if $[A,B]\neq 0$, the matrix elements of $A$ and $B$  in the eigenbasis of $C(g)$ have off-diagonal elements depending on $g$.}

Let us take a closer look at the situation for $p>2$. The condition $\[ D_1^G \,, D_2^G \]=0$ reduces to $\[ \cH_1 \,, \cH_2 \]=0$. If we impose this condition for any $\{ n_{ij} (\sigma) \}$, we get
\begin{align}
0 &= \sum_{ij,i'j'} n_{ij} (\sigma) \, n_{i'j'} (\sigma) \, [ \cH_{1, ij} \,, \cH_{2, i'j'} ]
\\
&= \sum_{ij} n_{ij} (\sigma)^2 \, [ \cH_{1, ij} \,, \cH_{2,ij} ]
+ \sum_{ijk} n_{ij} (\sigma) \, n_{ik} (\sigma) \( [ \cH_{1, ij} \,, \cH_{2,ik} ] + [ \cH_{1, ik} \,, \cH_{2,ij} ] \) 
\end{align}
giving us
\begin{equation}
0 = [ \cH_{1, ij} \,, \cH_{2,ij} ]
= [ \cH_{1, ij} \,, \cH_{2,ik} ] + [ \cH_{1, ik} \,, \cH_{2,ij} ].
\label{commutation cHij}
\end{equation}
By explicit computation, one finds
\begin{equation}
[ \cH_{1, ij} \,, \cH_{2,ij} ] = 0 \qquad {\rm but} \qquad
[ \cH_{1, ij} \,, \cH_{2,ik} ] + [ \cH_{1, ik} \,, \cH_{2,ij} ] \neq  0
\label{commutation cHij eval}
\end{equation}
and hence $\[ D_1^G \,, D_2^G \] \neq 0$ for $p>2$.
We will see in Section \ref{sec:all-loop} that this is a generic feature of effective $U(p)$ theory Hamiltonians under some ansatz, and {\it not} due to potentially missing terms in $D_2^G$\,.

\bigskip
This trouble can be solved in the following way.
In the displaced corners approximation, we truncate the Hilbert space to a fixed number of columns, then take the large $N_c$ limit. 
We obtained the Hamiltonians $D_\ell^G$ after the Hilbert space truncation, but without taking the limit. In fact, in the continuum limit we find
\begin{equation}
\[ D_1^G \,, D_2^G \] = 0
\label{commute D12}
\end{equation}
showing that taking the large $N_c$ limit is a necessary step.

Note that there is a caveat in the displaced corners approximation.
Even if we pick up an operator in the distant region $l_1 \gg l_2$\,, the difference $(l_1-l_2)$ keeps decreasing due to the operator mixing, until it hits the Young diagram constraints $l_1 \ge l_2$\,. 
The original one-loop mixing matrix no longer takes the simple form \eqref{Dil12 GGB} when two columns have comparable lengths. We expect that these boundary effects on the anomalous dimensions are negligible at the leading order of large $N_c$\,.

\subsection{Continuum limit}\label{sec:continuum limit}

We take the continuum limit following \cite{Carlson:2011hy,Koch:2011hb,deMelloKoch:2012sv}.

We begin with the ansatz for the dilatation eigenstates,
\begin{equation}
\cO_f (\sigma) = \sum_{l_1 , l_2 , \dots, l_p} \!\!\!\!{}^\prime \ 
f(l_1, l_2, \dots, l_p) \, O^{R, r_1} (\sigma), \qquad
N_c \ge l_1 \ge l_2 \ge \dots \ge l_p \ge 0, \qquad \sum_{i=1}^p l_i = n_Z 
\label{def:ansatz Of}
\end{equation}
where we specify the column lengths of $r_1$ by $(l_1, l_2 , \dots, l_p)$, and $\Sigma^\prime$ means the sum over $\{ l_i \}$ under the constraints shown in \eqref{def:ansatz Of}.
The action of the operators \eqref{def:dpmh} on $O_f$ can be written as
\begin{equation}
\begin{aligned}
d_i^- \, \cO_f (\sigma) &\simeq \sum_{l_1 , l_2 , \dots, l_p} \!\!\!\!{}^\prime \ 
\sqrt{ h(i, l_i) } \, f( \dots, l_i +1 , \dots ) \, O^{R, r_1} (\sigma)
\\
d_i^+ \, \cO_f (\sigma) &\simeq \sum_{l_1 , l_2 , \dots, l_p} \!\!\!\!{}^\prime \ 
\sqrt{ h(i, l_i+1) } \, f( \dots, l_i -1 , \dots ) \, O^{R, r_1} (\sigma)
\end{aligned}
\label{dpmh cont}
\end{equation}
where $\simeq$ means that we neglect potential contributions from the boundary of the summation range.

Consider the following large $N_c$ limit
\begin{equation}
n_Z \sim O(N_c), \qquad 
l_1 \sim O(N_c), \qquad
l_i \sim O( \sqrt{N_c} ) \qquad ( i =2,3, \dots, p )
\label{def:sqrt Nc limit}
\end{equation}
which is similar to the limit discussed in \cite{DeComarmond:2010ie}.
We prefer the square-root scaling $l_i \sim O(\sqrt{N_c})$ to the linear scaling $l_i \sim O(N_c)$, because the difference equations are rather trivial in the latter limit.
Physically, the system \eqref{def:sqrt Nc limit} consists of one nearly maximal, and $(p-1)$ far-from maximal spherical giant gravitons. 
The constraint $\sum_i l_i = n_Z$ becomes somewhat trivial because $n_Z \sim l_1$\,.

We introduce the rescaled variables and functions
\begin{gather}
y_i = \frac{l_i }{\sqrt{\alpha N_c }} \,, \qquad
\sqrt{ \frac{N_c}{\alpha} } \, \ge y_1 \ge y_2 \ge \dots \ge y_p \ge  0
\label{def:yi} \\
F (y_1, y_2, \dots, y_p) \equiv 
f \( \frac{l_1 }{\sqrt{\alpha N_c }} \,, \frac{l_2 }{\sqrt{\alpha N_c }} \,, \dots \,, \frac{l_p }{\sqrt{\alpha N_c }} \) .
\end{gather}
We keep $y_1$ to simplify our notation, even though $y_1 = O(\sqrt{N_c}) \gg 1$.
It follows that
\begin{equation}
\begin{aligned}
h (i, l_i) &= N_c + i - y_i \sqrt{ \alpha N_c} 
\\
f( \dots, l_i \pm 1 , \dots ) &= F \( \dots, y_i \pm \frac{1}{\sqrt{\alpha N_c }} \,, \dots \) .
\end{aligned}
\end{equation}
In the continuum limit, the difference operators $\cH_{1,ij} \,, \cH_{2,ij}$ in \eqref{def:cH1}, \eqref{def:cH2} become
\begin{equation}
\cH_{1,ij} \to \cD_{ij} , \qquad
\cH_{2,ij} \to 2 N_c \, \cD_{ij} \,, \qquad
\cD_{ij} \equiv \frac{\alpha}{4} \, y_{ij}^2 - \frac{1}{\alpha} \, \frac{\partial^2}{\partial y_{ij}^2}
\label{def:cDij}
\end{equation}
where $y_{ij} = y_i - y_j$\,.
This suggests that the one-loop and two-loop dilatations commute in this limit,
\begin{equation}
\[ D_1^G \,, D_2^G \] \to 0 .
\end{equation}
The spectrum of $D_1^G$ is discussed in detail in Appendix \ref{app:explicit spectrum}.

\section{All-loop ansatz}\label{sec:all-loop}

We conjecture that perturbative dilatation operators at all loops in the continuum limit \eqref{def:sqrt Nc limit} takes the form
\begin{equation}
D^G = D_0 + f_c (\lambda) \, \sum_{\substack{i,j=1 \\ i \neq j}}^p n_{ij} ( \sigma ) \, \cD_{ij} \,, \qquad
\lambda \equiv N_c \, g_{\rm YM}^2 \,.
\label{DG all-loop}
\end{equation}
A related argument was given in \cite{Koch:2013xaa}, where they showed that the mixing of $Y$'s at higher loops takes the same form as the one-loop mixing.

\subsection{Constraints on higher-loop dilatations}

We expand the dilatation operator on the Gauss graph basis at weak coupling as
\begin{equation}
D^G (g_{\rm YM}) = \sum_{\ell =0}^\infty \( \frac{g_{\rm YM}}{4 \pi} \)^{2 \ell} D_\ell^G 
\label{def:D^G any g}
\end{equation}
and make the following ansatz for $D_\ell^G$ in the leading order of large $N_c$\,,
\begin{equation}
D_\ell^G = \sum_{k=1}^\ell N_c^{\ell-k} x_k \, \cH_k , \qquad
\cH_\ell = \sum_{i \neq j}^p n_{ij} (\sigma) \, \cH_{\ell, ij} 
\label{ansatz DG-loop}
\end{equation}
where $\{ x_k \}$ are numerical constants of $O(1)$.
The first equation \eqref{ansatz DG-loop} means that the $\ell$-loop dilatation contains lower-loop difference operators multiplied by powers of $N_c$\,.
The second equation means that $\cH_\ell$ depends only on a pair of column labels $(i,j)$ coupled to $n_{ij} (\sigma)$. We impose this condition because $n_{ij} (\sigma)$ should count the number of open string modes stretching between the $i$-th and $j$-th giant graviton brane.

We further assume that
\begin{equation}
\cH_{\ell, ij} \equiv \sum_{m} \tilde x_{\ell m} \, 
\cP_{\ell, m} (d_i^\dagger , d_j^\dagger , d_i, d_j), \qquad
\cP_{\ell, m} \, \ \text{contains $\ell$\ $d^\dagger$'s followed by $\ell$\ $d$'s}
\label{ansatz Pij-loop}
\end{equation}
where $\{ \tilde x_{\ell, m} \}$ are numerical constants of $O(1)$, and $\cP_{\ell, m}$ is a polynomial of the difference operators.
The form of $\cP_{\ell, m}$ originates from the perturbative dilatation operators of $\cN=4$ SYM discussed in Section \ref{sec:pert dil}.
It is known that there is a correspondence between the terms of $\Delta^{(2)}_{ij}$ in \eqref{two-loop Delta ij} and those of $D_2$\,, according to the two-loop computation \cite{deMelloKoch:2012sv}.\footnote{For example, $(d_i^\dagger)^2 d_j^2$ removes two boxes from the $j$-th column and add two boxes to the $i$-th column. This term comes from $(\Delta_{ij}^+)^2$ which roughly corresponds to $\Tr (ZZW \check Z \check Z \check W)$.}
Since the $\ell$-loop dilatation operator $D_\ell$ should remove at most $\ell$ fields and add $\ell$ fields, we arrive at the ansatz of $\cP_{\ell, m}$ in \eqref{ansatz Pij-loop}.\footnote{We can also explain the powers of $N_c$ in \eqref{ansatz DG-loop} from the fact that $D_\ell$ removes $\ell$ fields and adds $\ell$ fields.}

Let us revisit the commutation relations in Section \ref{sec:commutation}. Now we impose
\begin{equation}
[D^G (g_{\rm YM}) , D^G (g'_{\rm YM})] = 0, \quad (\forall g_{\rm YM}, g'_{\rm YM} )
\qquad \Leftrightarrow \qquad
[D_k^G, D_l^G]=0, \quad (\forall k, l )
\label{Dkl commute}
\end{equation}
by generalizing \eqref{commute D12}. This is a crucial assumption in our discussion, and should be justified in future.
The equation \eqref{Dkl commute} implies that the mixing problem can be ``one-loop exact'' in the sense that the eigenvectors remain unchanged at higher loop orders.\footnote{The author thanks an anonymous referee of JHEP for correcting mistakes in the previous version and emphasizing this point.}

The first equation of \eqref{commutation cHij} is generalized to
\begin{equation}
[ \cH_{\ell, ij} \,, \cH_{\ell',ij} ] = 0 \qquad (\forall \ell, \ell').
\label{commute ij-loop}
\end{equation}
It is straightforward to enumerate all possible solutions of \eqref{commute ij-loop}, or equivalently the general form of $\cP_{\ell, m}$\,, with the help of {\tt Mathematica}.

It turns out that at $\ell$-loop, there are $(\ell+1)$ independent solutions of the equation \eqref{commute ij-loop}.
At one-loop, there are two solutions
\begin{alignat}{9}
\cP_{ij,10} &= d_i^\dagger d_j + d_j^\dagger d_i  & &\equiv \cJ_{ij} 
\label{def:cJij}
\\[1mm]
\cP_{ij,11} &= d_{ij}^\dagger \, d_{ij} & &= \cH_{1, ij}
\end{alignat}
where $\cH_{1, ij}$ is given in \eqref{def:cH1} and
\begin{equation}
d_{ij}^\dagger = d_i^\dagger - d_j^\dagger \,, \qquad
d_{ij} = d_i - d_j \,.
\end{equation}
We also define
\begin{equation}
\cI_{ij} \equiv d_i^\dagger d_i + d_j^\dagger d_j = d_{ij}^\dagger \, d_{ij} - \cJ_{ij}
\label{def:cIij}
\end{equation}
which satisfies
\begin{equation}
\[ \cI_{ij} \,, \cJ_{ij} \] = 0.
\label{commute IJ}
\end{equation}
At two-loop, there are three solutions,
\begin{alignat}{9}
\cP_{ij,20} &= (d_i^\dagger )^2 (d_j)^2 + 2 \, d_i^\dagger d_j^\dagger d_i d_j 
+ (d_j^\dagger )^2 (d_i)^2 
& &= \cJ_{ij}^{(2)}
\\[1mm]
\cP_{ij,21} &= d_{ij}^\dagger \, \cJ_{ij} \, d_{ij}
& &= \cH_{2, ij} 
\\[1mm]
\cP_{ij,22} &=  (d_{ij}^\dagger)^2 \, (d_{ij})^2 
& &= \  : \cH_{1, ij}^2 \, :
\end{alignat}
where $\cH_{2, ij}$ is given in \eqref{def:cH2} and
\begin{equation}
\cJ_{ij}^{(n)} \equiv \ : \! (\cJ_{ij})^n \! : \ 
= \sum_{m=0}^n \binom{n}{m} (d_i^\dagger)^m (d_j^\dagger)^{n-m} \, (d_i)^{n-m} (d_j)^m .
\label{def:varpin}
\end{equation}

At higher loops, we find that all solutions at $\ell$-loop can be written in the form
\begin{equation}
\cQ_{ab} = \cQ_{ab,ij} \equiv ( d_{ij}^\dagger )^a \, \cJ_{ij}^{(b)} \, ( d_{ij} )^a  \,,
\qquad (a = \ell - b = 0, 1, \dots, \ell).
\label{def:Qabij}
\end{equation}
We checked that no more solutions exist up to four-loop. 
The two-parameter family of difference operators \eqref{def:Qabij} mutually commute,
\begin{equation}
\[ \cQ_{ab} \,, \cQ_{a'b'} \] = 0 \qquad (\forall a,b,a',b')
\label{commute Qabij}
\end{equation}
which follows from \eqref{def:cIij} and \eqref{commute IJ}.
Our ansatz for the $\ell$-loop dilatation in \eqref{ansatz DG-loop} becomes
\begin{equation}
\cH_{\ell, ij} = \sum_{m=0}^\ell \tilde x_{\ell m} \, \cQ_{\ell-m, m} \,.
\label{ansatz DG-loop-2}
\end{equation}

\bigskip
The second equation of \eqref{commutation cHij} generalized to higher loops reads
\begin{equation}
[ \cH_{\ell, ij} \,, \cH_{\ell',ik} ] + [ \cH_{\ell, ik} \,, \cH_{\ell',ij} ] = 0 \qquad (\forall \ell, \ell').
\label{commute ij-loop-2}
\end{equation}
Some of $\cQ_{ab,ij}$ in \eqref{def:Qabij} up to two loops satisfy these conditions. 
At three-loops, no linear combination of $\( \cQ_{3,0} \,, \cQ_{2,1} \,, \cQ_{1,2} \,, \cQ_{0,3} \)$ satisfy \eqref{commute ij-loop-2} against the two-loop dilatation.
Thus, we should trust the discrete form of our all-loop ansatz \eqref{ansatz DG-loop} only at $p=2$.

This conclusion is not surprising.
The effective Hamiltonian $\cH_{\ell, ij}$ is a linear combination of $\cQ_{ab,ij}$ in \eqref{def:Qabij}, which are polynomials of $\cJ_{ij}$\,.
However, $\cJ_{12}$ and $\cJ_{23}$ as the $\alg{su}(p)$ generators \eqref{GL(p) commutation} do not commute:
\begin{equation}
\[ \cJ_{12} \,, \cJ_{23} \] \sim
\[ 
\begin{pmatrix}
0 & 1 & 0\\
1 & 0 & 0 \\
0 & 0 & 0
\end{pmatrix} , 
\begin{pmatrix}
0 & 0 & 0\\
0 & 0 & 1 \\
0 & 1 & 0
\end{pmatrix} \] =
\begin{pmatrix}
0 & 0 & 1\\
0 & 0 & 0\\
-1&0 & 0
\end{pmatrix}
\neq 0
\end{equation}
which makes it hard to solve \eqref{commute ij-loop-2}.
See also the discussion in Section \ref{sec:commutation}.

\bigskip
We can determine the numerical coefficients in \eqref{ansatz DG-loop}, \eqref{ansatz DG-loop-2} up to two loops. By comparing $\cQ_{ab}$ with the perturbative results \eqref{Del GGB-cH}, we find
\begin{equation}
\begin{aligned}
D_1^G &= - \sum_{i \neq j} n_{ij} (\sigma) \, \pare{ 0 \cdot \cQ_{0,1} + 1 \cdot \cQ_{1,0} }
\\
D_2^G &= - \sum_{i \neq j} n_{ij} (\sigma) \, \pare{ 
0 \cdot \cQ_{0,1} + \( L - 2 N_c \) \cQ_{1,0} 
+ 0 \cdot \cQ_{0,2} + 1 \cdot \cQ_{1,1} + 0 \cdot \cQ_{2,0} } 
\end{aligned}
\label{DlG coefficients}
\end{equation}
implying that most coefficients vanish in $\cN=4$ SYM.
This result is also consistent with our assumption in \eqref{ansatz DG-loop} that $\{ \tilde x_{\ell m} \}$ are numerical constants of $O(1)$.

\subsection{Continuum limit at higher loops}\label{sec:cont higher}

By combining \eqref{ansatz DG-loop} and \eqref{ansatz DG-loop-2}, we obtain a conjecture of the $\ell$-loop dilatation operator,
\begin{equation}
D_\ell^G = \sum_{i \neq j}^p n_{ij} (\sigma) \, 
\sum_{k=1}^\ell \sum_{m=0}^k N_c^{\ell-k} \, C_{k,m} \, \cQ_{k-m, m} \,, \qquad
C_{k,m} = x_k \, \tilde x_{k m}  = O(N_c^0).
\label{conjecture DlG}
\end{equation}
The terms with $k < \ell$ are part of the dilatation at lower loop orders, combined with powers of $N_c$\,.

Let us take the continuum limit \eqref{def:sqrt Nc limit}. The commuting operators $\cQ_{ab}$ in \eqref{def:Qabij} scale as\footnote{Before the continuum limit, $\cQ_{ab}$ scales as $N_c^{a+b}$.}
\begin{equation}
\cQ_{\ell-m, m} \sim N_c^m 
\label{cQlm scaling-1}
\end{equation}
and in particular
\begin{equation}
\begin{aligned}
\cQ_{0,m} &= (2 N_c)^m + O(N_c^{m-1/2}) 
\\[1mm]
\cQ_{1,m} &= (2 N_c)^m \, \cD_{ij} + O(N_c^{m-1/2}) 
\\[1mm]
\cQ_{2,m} &= (2 N_c)^m  \, \(  \frac{\alpha^2}{16} \, y_{ij}^4
- \frac{y_{ij}^2}{2} \, \frac{\partial^2}{\partial y_{ij}^2}
+ \frac{1}{\alpha^2} \, \frac{\partial^4}{\partial y_{ij}^4} \) + O(N_c^{m-1/2})
\\[1mm]
\cQ_{3,m} &= (2 N_c)^m \, \(  \frac{\alpha^3}{64} \, y_{ij}^6
- \frac{3 \, y_{ij}^4}{16} \, \frac{\partial^2}{\partial y_{ij}^2}
+ \frac{3 \, y_{ij}^2}{4} \, \frac{\partial^4}{\partial y_{ij}^4}
- \frac{1}{\alpha^2} \, \frac{\partial^6}{\partial y_{ij}^6} \) + O(N_c^{m-1/2})
\end{aligned}
\label{cQim continuum}
\end{equation}
where $\cD_{ij}$ is given in \eqref{def:cDij}. From this observation, we can refine \eqref{cQlm scaling-1} as
\begin{equation}
\cQ_{\ell, m} = (2 N_c)^m \, : \cD_{ij}^{\ell} : + O(N_c^{m-1/2})
\label{cQlm scaling-2}
\end{equation}
where $: \cD_{ij}^{\ell} :$ means that the derivative $(\partial/\partial y_{ij})$ should not hit $y_{ij}$ in the subsequent $\cD_{ij}$'s.

Then, our conjectured dilatation operator \eqref{conjecture DlG}  becomes
\begin{equation}
D_\ell^G = \sum_{i \neq j}^p n_{ij} (\sigma) \, 
\sum_{k=1}^\ell \sum_{m=0}^k N_c^{\ell-k+m} \, (2^m C_{k,m}) \,: \cD_{ij}^{k-m} :   \,.
\label{conjecture DlG-2}
\end{equation}
We find that the terms $m=k$ are leading at large $N_c$\,. However, perturbative data \eqref{DlG coefficients} shows that $C_{k,k}=0$. The terms $m=k-1$ gives the first non-vanishing term, which is proportional to the one-loop result $D_1^G$. The terms $m \le k-2$ are negligible as long as $C_{k,m} = O(N_c^0)$.

\bigskip
Given \eqref{conjecture DlG-2} we can formally sum up the perturbation series,
\begin{equation}
D^G = D_0 + \sum_{\ell=1}^\infty g_{\rm YM}^{2 \ell} \, D_\ell^G
\ \to \ 
D_0 + N_c^{-1} f_c (\lambda) \sum_{i \neq j} n_{ij} (\sigma) \, \cD_{ij}
\label{conjecture DG-all}
\end{equation}
where $\lambda = N_c \, g_{\rm YM}^{2}$ is the 't Hooft coupling and
\begin{equation}
f_c (\lambda) 
= \sum_{\ell=1}^\infty \lambda^{\ell} \, \sum_{k=1}^{\ell} 2^{k-1} \, C_{k,k-1} \,.
\end{equation}
This is the result quoted in \eqref{DG all-loop}.
According to Appendix \ref{sec:cont case}, the operator $\cD \equiv \sum n_{ij} (\sigma) \, \cD_{ij}$ has the eigenvalues \eqref{continuous Sch ev}. Thus
\begin{equation}
D^G - D_0 = N_c^{-1} f_c (\lambda) \pare{ 
n_Y + 2 \sum_{a=1}^{p-1} \( 2m_a + 1\) \tilde \lambda_a ( \{ n_{ij} \} ) }
\label{conjecture DG-all-2}
\end{equation}
where $\lambda_a$ depends on $n_{ij} (\sigma)$ and has the same order as $n_Y$\,.
We will see in Appendix \ref{sec:disc case} that the non-negative integers $m_a$ should be bounded from above, and at most $O(N_c)$.
Neglecting $O(1)$ quantities, the equation \eqref{conjecture DG-all-2} becomes
\begin{equation}
D^G \ \to \ 
D_0 + \tilde f (\lambda) \sum_{a=1}^{p-1} \frac{m_a}{N_c} \, \tilde \lambda_a ( \{ n_{ij} \} ) 
\label{DG cont all-loop}
\end{equation}
with $\tilde f (\lambda) = 4 f_c (\lambda)$. When $p=2$, we obtain
\begin{equation}
\begin{gathered}
D^G = L + \tilde f (\lambda) \frac{m}{N_c} \, n_{12} (\sigma), \quad
\tilde f (\lambda) = \frac{\lambda}{2 \pi^2} + O(\lambda^2)
\\
n_{12} (\sigma) \in \bb{Z}_{\ge 0} \,, \qquad
m = 1,2, \dots, \left\lceil N_c - \frac{n_Z}{2} + 1 \right\rceil 
\end{gathered}
\label{DG cont all-loop p=2}
\end{equation}
where we used \eqref{def:DgYM} and \eqref{app:D1G p=2}.

The factor $m_a/N_c$ in \eqref{DG cont all-loop} has the spacing of order $1/N_c$\,, which becomes continuous at large $N_c$\,.
If $f_c(\lambda)$ remains non-zero at $\lambda \gg 1$, then the above ansatz should describe (semi)classical motion of the system with D-branes and strings.
Importantly, this excitation spectrum should be gapless.

\section{Strong coupling}\label{sec:strong coupling}

We want to reproduce the dilatation spectrum \eqref{DG cont all-loop} at strong coupling.
Since the energy of excited states is continuously connected to the BPS state, we take a classical D3-brane action and study the solution around the BPS giant graviton.

\subsection{D3-brane action}\label{sec:classical D3 action}

The action for a single D3-brane is given by
\begin{equation}
S = S_{\rm DBI} + S_{\rm CS}
= -T_3 \int_{\Sigma_4} d^4 \xi \, e^{-\varphi}\sqrt{- \det\(
G_{ab} + B_{ab} + 2 \pi \alpha' F_{ab} \)}
+ T_3 \int_{\Sigma_4} C^{(4)}
\label{def:D3 action}
\end{equation}
where $\Sigma_4$ is the worldvolume, $G_{ab} = g_{\mu\nu} \, \partial_a X^\mu \partial_b X^\nu$ is the induced metric.
We consider a D3-brane wrapping S$^3$ inside \AdSxS,
\begin{equation}
{\rm D3} \ : \ \Sigma_4 \ \to \ 
\bb{R} \times {\rm S}^3 \ \subset \ 
{\rm AdS}_5 \times {\rm S}^5 
\label{def:Sigma4}
\end{equation}
which includes the spherical giant graviton.
There is no $B_{ab}$ and $F_{ab}$ in the background, and the dilaton is constant, $e^{\varphi} =g_s$\,.
The constant $T_3$ is given by \cite{McGreevy:2000cw}
\begin{equation}
T_3 = \frac{2 \pi}{g_s (4 \pi^2 \alpha')^2}
= \frac{N_c}{R^4 \, \Omega_3}
\label{def:T3}
\end{equation}
where $R$ is the radius of \AdSxS\ and $\Omega_3 = 2 \pi^2$ is the volume of S$^3$ with the unit radius.
Our notation for the \AdSxS\ geometry is explained in Appendix \ref{app:geometry}.

The induced metric is written as
\begin{multline}
G_{ab} = R^2 \Bigl\{ - \partial_a t \, \partial_b t 
+ \frac{\partial_a \rho \, \partial_b \rho}{4 (\rho -1) \rho^2}
+ \frac{\(\rho -1 \) \partial_a \phi \, \partial_b \phi}{\rho}
\\
+ \frac{\partial_a \eta \, \partial_b \eta 
+ \cos^2 \eta \, \partial_a \theta_1 \, \partial_b \theta_1 
+ \sin^2 \eta \, \partial_a \theta_2 \, \partial_b \theta_2 }{\rho} 
\Bigr\} .
\label{Gab eval}
\end{multline}
We choose the static gauge
\begin{equation}
t = \xi^0, \qquad
\theta_1 = \xi^1 \,, \qquad
\theta_2 = \xi^2 \,, \qquad
\eta = \xi^3
\label{sec:static gauge}
\end{equation}
and assume the ansatz
\begin{equation}
\rho = \rho (t, \eta), \qquad
\phi = \phi (t, \eta).
\label{GGG ansatz}
\end{equation}
The D3-brane action effectively becomes two-dimensional,\footnote{We chose the CS coupling so that the ground state satisfies $E=J$ including the sign, which can be flipped by $\phi \to -\phi$. The physical brane tension is proportional to $N_c/g_s$\,.}
\begin{equation}
S \equiv  \int_{\bb{R} \times {\rm S}^3} d^4 \xi \, \cL
= \frac{N_c}{g_s \, \Omega_3} \int d^4 \xi
\Biggl( - \sqrt{- \det G} + \delta_{a0} \, \sin \eta \cos \eta \, \frac{(\partial_a \phi)}{\rho^2} 
\Biggr) .
\label{D3 action 2d}
\end{equation}
The conserved charges can be computed in the standard way,
\begin{alignat}{9}
J &= \int_0^{\pi/2} d\eta \, {\tt j}^0 (t, \eta)
& &= \int d\eta \, \frac{\delta S}{\delta \partial_0 \phi}
\notag \\
E &= \int_0^{\pi/2} d \eta \, {\tt h} (t, \eta)
& &= \int d\eta \pare{ 
\sum_{X=\rho, \phi} \partial_a X \, \frac{\delta S}{\delta \partial_a X} - \cL
} .
\label{conserved charges}
\end{alignat}

\subsection{Classical solutions}

We study the effective two-dimensional action \eqref{D3 action 2d} around the spherical giant graviton solution as follows. We assume the static gauge, introduce the deformation parameter $\epsilon$, and solve the equations of motion (EoM) as a formal series of small $\epsilon$. 
The linearized EoM are given by a set of homogeneous partial differential equations, whose coefficients may depend on $\eta$. We remove the $\eta$ dependence by the separation of variables for the deformed degrees of freedom.

This procedure looks similar to the analysis of one-loop fluctuation \cite{Das:2000st,Sadri:2003mx}. 
Generally, however, not all off-shell fluctuations become the deformed solutions of the classical equations of motion.

Other deformations of the spherical giant graviton solution might be possible if the ansatz and the gauge choice are generalized.\footnote{The gauge choice may change the CS term.}
One can try to study the deformation in the AdS directions, and to search for the solutions with non-zero $U(1)$ field strength or other components of $SO(6)$ angular momenta.

\subsubsection{Ground state}

The ansatz for the ground (or BPS) state of a spherical giant graviton is \cite{McGreevy:2000cw} 
\begin{equation}
\rho = \text{constant} , \qquad
\phi = t .
\label{vacuum GG ansatz}
\end{equation}
The energy as a function of $\rho$ has a local minimum at $\rho = N_c/(g_s J)$, and the minimum value is
\begin{equation}
E = J .
\end{equation}

\subsubsection{Excited states}

We are interested in the non-BPS states which are continuously connected to the BPS state.
Let us generalize the ansatz by expanding around the ground state solution as
\begin{equation}
\rho = \frac{1}{j} + \epsilon \, \rho_1 (t, \eta), \qquad
\phi = t + \epsilon \, \phi_1 (t, \eta) , \qquad
j \equiv \frac{g_s J}{N_c} \,.
\label{rhophi epsilon ansatz}
\end{equation}

We consider the EoM  for three cases, $j=0$, $0<j<1$ and $j=1$. No non-trivial solutions are found for the cases $j=0,1$, as discussed in Appendix \ref{app:special sol}. 
When $0<j<1$, the EoM for $\phi$ and $\rho$ take the form
\begin{align}
\frac{j^2 \partial_t \rho_1}{j-1}
+ \partial_t^2 \phi_1
&=
\partial_\eta^2 \phi _1
+2 \cot (2 \eta ) \partial_\eta \phi_1
\label{eom-phi1}
\\[1mm]
- \frac{4 (j-1)\partial_t \phi _1}{j^2}
+ \partial_t^2 \rho_1
&=
\partial_\eta^2 \rho _1
+2 \cot (2 \eta ) \partial_\eta \rho_1 
\label{eom-rho1}
\end{align}
which can be solved by separation of variables.
The RHS of \eqref{eom-phi1}, \eqref{eom-rho1} are identical to the Laplacian on S$^3$, whose normalizable solutions are given by the spherical harmonics \eqref{app:spherical harmonics S3}. 
Since our $\phi_1$ and $\rho_1$ are independent of $\theta_1 \,, \theta_2$\,, we set
\begin{equation}
\rho_1 (t, \eta) = \tilde \rho_1 (t) \, \Phi_{k,0,0} (\eta) , \qquad
\phi_1 (t, \eta) = \tilde \phi_1 (t) \, \Phi_{k,0,0} (\eta) .
\label{ansatz rho-phi1}
\end{equation}
Then the general solution of the equations \eqref{eom-phi1}, \eqref{eom-rho1} is given by
\begin{align}
\tilde \rho_1 (t) &= \frac{1}{j^2} \Biggl[
c_{(1)} \, \Bigl( (k+2) \cos (k t) + k \cos ((k+2) t) \Bigr)
+ c_{(2)} \, \cos t \, \sin ((k+1) t)
\notag \\
&\hspace{10mm}
+ c_{(3)} \, \Bigl( k \sin ((k+2) t) - (k+2) \sin (k t)  \Bigr)
+ c_{(4)} \, \sin t \, \sin ((k+1) t)
\Biggr]
\label{sol rho1} \\
\tilde \phi_1 (t) &= \frac{1}{2 (j-1)} \Biggl[
c_{(1)} \, \Bigl( (k+2) \sin (k t) - k \sin ((k+2) t) \Bigr)
- c_{(2)} \, \sin t \, \sin ((k+1) t)
\notag \\
&\hspace{10mm}
+ c_{(3)} \, \Bigl( (k+2) \cos (k t) + k \cos ((k+2) t) \Bigr)
+ c_{(4)} \, \cos t \, \sin ((k+1) t)
\Biggr]
\label{sol phi1}
\end{align}
where $c_{(i)} \ (i=1,2,3,4)$ are integration constants.

Let us compute the corrections to the conserved charges from \eqref{ansatz rho-phi1},
\begin{equation}
E = \sum_{n=0} \epsilon^n \, E_{(n)} = \sum_{n=0} \epsilon^n \, \int_0^{\pi/2} d \eta \, {\tt h}_{(n)} \,,
\qquad
J = \sum_{n=0} \epsilon^n \, J_{(n)} = \sum_{n=0} \epsilon^n \, \int_0^{\pi/2} d \eta \, {\tt j}_{(n)}^0 \,.
\end{equation}
It follows that
\begin{equation}
{\tt h}_{(1)} = {\tt j}_{(1)}^0 =
\frac{N_c}{g_s \, \Omega_3} \, \sin \eta  \cos \eta \ \Phi_{k,0,0} (\eta) \Bigl\{
(1-j) \partial_t \tilde \phi_1 - j^2 \tilde \rho_1 (t) \Bigr\} .
\end{equation}
Only the $k=0$ term remains non-zero after the integration over S$^3$ owing to the orthogonality \eqref{ON Sph Harm}. We remove the first-order correction to the conserved charges by setting
\begin{equation}
E_{(1)} = J_{(1)} \ \propto \ (4 c_{(1)} + c_{(4)}) = 0.
\end{equation}
The difference $(E-J)$ is non-zero at the second order in the $\epsilon$ expansion,
\begin{align}
{\tt h}_{(2)} &- {\tt j}_{(2)}^0 
= \frac{N_c}{g_s \, \Omega_3} \, \frac{\sin 2 \eta}{16 (1-j)} \, \Bigl\{
\Phi_{k,0,0} (\eta)^2 \left( j^4 (\partial_t \tilde \rho_1)^2 
+ 4 (j-1)^2 (\partial_t \tilde \phi_1)^2 \right)
\notag \\[1mm]
&\hspace{70mm}
- (\partial_\eta \Phi_{k,0,0})^2 \left( j^4 \, \tilde \rho_1^2 
+4 (j-1)^2 \, \tilde \phi_1^2 \right)
\Bigr\} 
\label{E-J 2nd order} \\[1mm]
&\simeq \frac{N_c}{g_s \, \Omega_3} \, \frac{\sin 2 \eta \, \Phi_{k,0,0} (\eta)^2}{16 (1-j)} \, \Bigl\{
\left( j^4 (\partial_t \tilde \rho_1)^2 
+ 4 (j-1)^2 (\partial_t \tilde \phi_1)^2 \right)
- k (k+2) \left( j^4 \, \tilde \rho_1^2 
+4 (j-1)^2 \, \tilde \phi_1^2 \right)
\Bigr\} 
\notag
\end{align}
where $\simeq$ denotes the equality after the integration over S$^3$ coming from \eqref{identity int SpH}.

Here we encounter apparent inconsistency.
The corrections to the conserved charges \eqref{E-J 2nd order} may depend on $t$ even after the integration over S$^3$.
This is partly because our ansatz \eqref{rhophi epsilon ansatz} solves the EoM only at $O(\epsilon)$ whereas the corrections are $O(\epsilon^2)$.
This explanation is not entirely correct because the solutions at $O(\epsilon^2)$ do not seem to change $E-J$ at $O(\epsilon^2)$.
Fortunately we can remove the $t$ dependence either by adjusting the constants $\{ c_i \}$, or by setting $k=0$.

The general $k>0$ solutions are given by
\begin{equation}
\begin{aligned}
\rho_1 (t, \eta) &= \frac{c_k (k+1) }{j^2 (k+2)} \, \sin ((k+2) t) \, \Phi_{k,0,0} (\eta )
\\[1mm]
\phi_1 (t, \eta) &= \frac{c_k (k+1) }{2 (j-1) (k+2)} \, \cos ((k+2) t) \, \Phi_{k,0,0} (\eta )
\end{aligned}
\label{k>0 sol ep1}
\end{equation}
which has the dispersion relation
\begin{equation}
E - J = \frac{N_c}{g_s} \, \frac{\epsilon ^2 \, c_k^2 \, (k+1)^2}{8 (1-j) (k+2)} \,.
\end{equation}
Here $k$ should be a positive even integer as in Appendix \ref{app:sph harm}, and $c_k$ should remain finite as $k \gg 1$ in order to keep \eqref{k>0 sol ep1} finite.
The general $k=0$ solution is
\begin{equation}
\rho_1 (t, \eta) = \frac{c_{(2)} \sin (2 t)-c_{(4)} \cos (2 t)}{2 \sqrt{2} \pi j^2} \,,\qquad
\phi_1 (t, \eta) = \frac{(4 c_{(3)} - c_{(2)}) +c_{(2)} \cos (2 t) + c_{(4)} \sin (2 t ) }{4 \sqrt{2} \pi  (j-1)} 
\label{k=0 sol ep1}
\end{equation}
with the dispersion relation
\begin{equation}
E - J = \frac{N_c}{g_s} \, \frac{\epsilon ^2 \( c_{(2)}^2 + c_{(4)}^2 \)}{32 \pi^2 (1-j)} \,.
\end{equation}
In \eqref{k=0 sol ep1}, the parameter $c_{(3)}$ is redundant because it just shifts the origin of $\phi$. Also, the terms proportional to $c_{(2)}$ coincide with the $k=0$ case of the previous solution \eqref{k>0 sol ep1}.

\vskip 10mm

\begin{figure}[H]
\begin{center}
\includegraphics[scale=0.9]{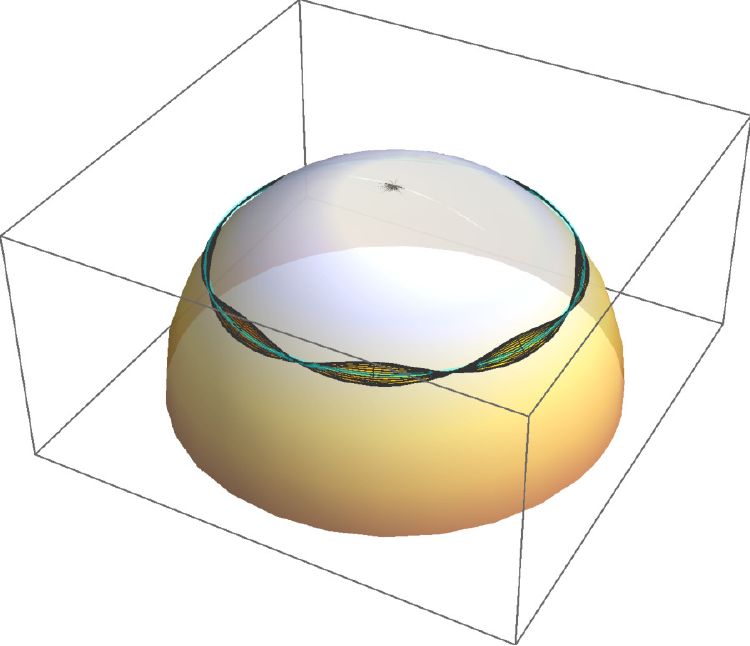}
\hspace{3mm}
\includegraphics[scale=0.9]{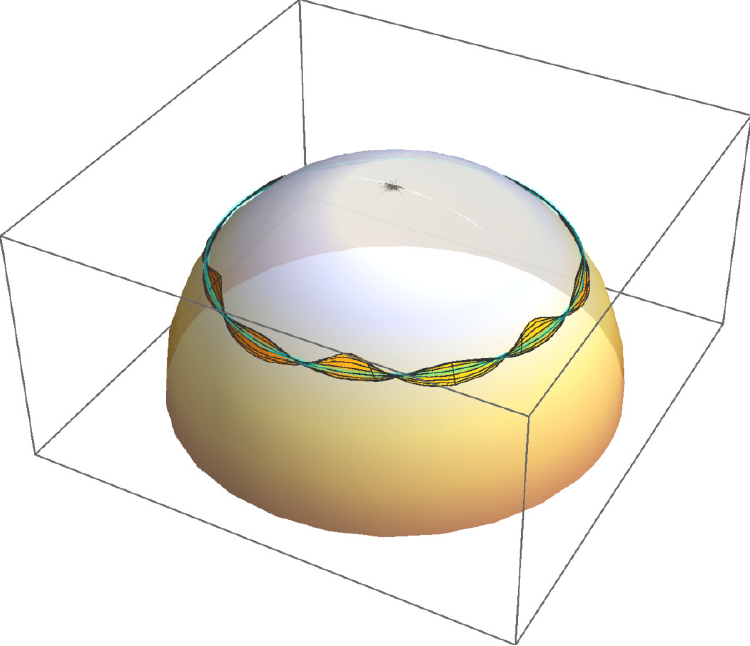}

\vskip 3mm
\includegraphics[scale=0.9]{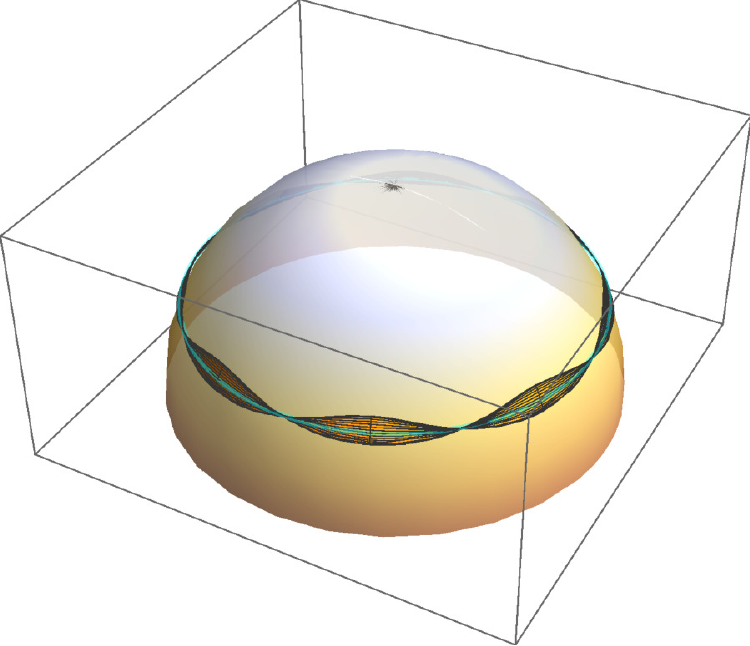}
\hspace{3mm}
\includegraphics[scale=0.9]{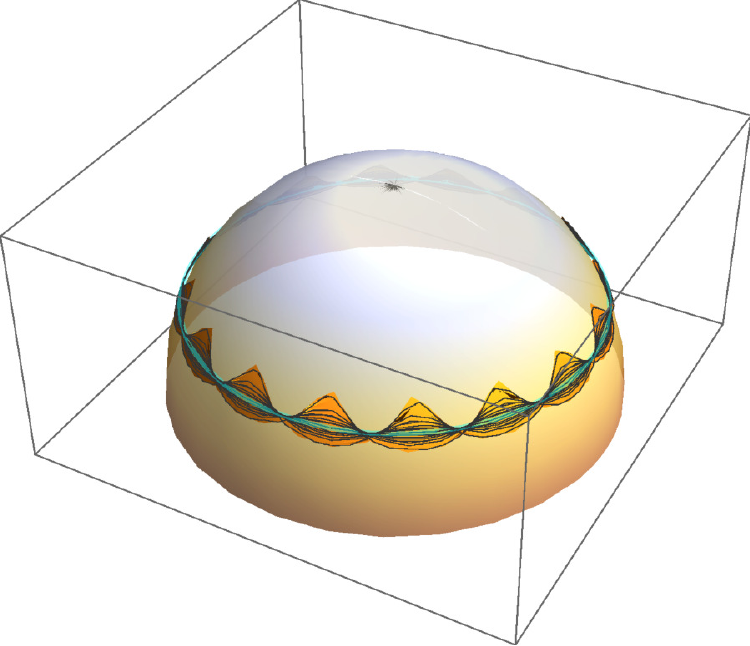}

\caption{The trajectory of oscillating giant gravitons with KK modes. The green lines denote the BPS ground states. We set $(j,k)=(0.4,2)$ for top left, $(j,k)=(0,4,4)$ for top right, $(j,k)=(0.2,2)$ for bottom left, $(j,k)=(0,2,6)$ for bottom right. Other parameters are $R=1, \epsilon=0.15, c_k=1$.}
\label{fig:oscillating giants KK}
\end{center}
\end{figure}

\begin{figure}[h]
\begin{center}
\includegraphics[scale=0.9]{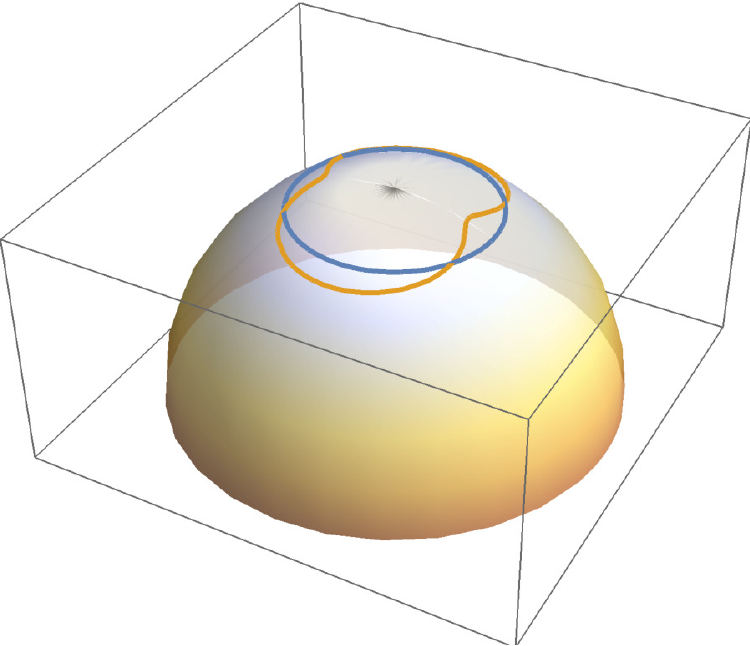}
\hspace{3mm}
\includegraphics[scale=0.9]{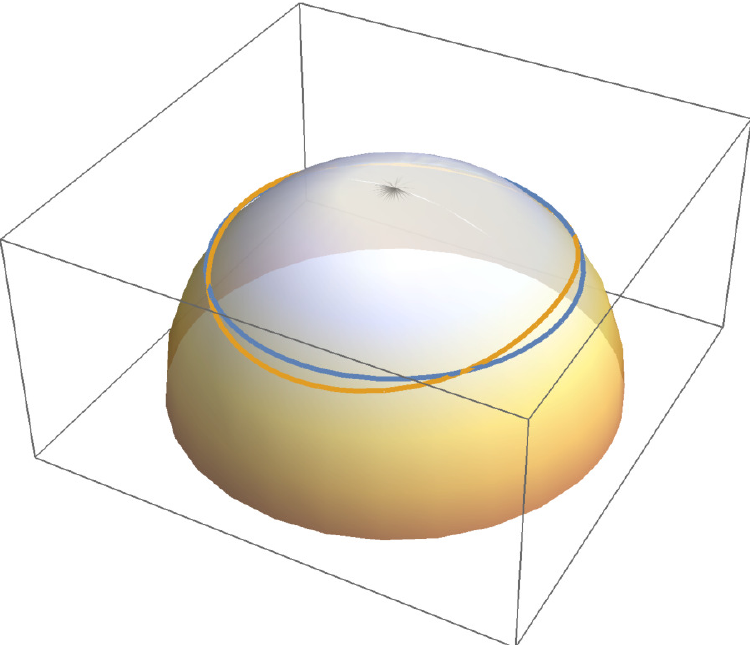}

\caption{Giant gravitons at the vacuum (blue) and excited (orange) states. We put $j=0.8$ for the left and $j=0.4$ for the right figure. Other parameters are $R=1, \epsilon=0.05, {\tt c}=1, {\tt d}=0$.}
\label{fig:oscillating giants 0}
\end{center}
\end{figure}

The profile of those with KK modes is depicted in Figure \ref{fig:oscillating giants KK}, and the profile of the oscillating giant gravitons without KK modes is in Figure \ref{fig:oscillating giants 0}.
The former solutions expand and shrink over the area of $O(\epsilon^2)$.
The latter solutions are point-like and oscillate around the BPS configuration.

\subsubsection{Excited states at higher orders}

We can take a linear combination of the solutions at $O(\epsilon)$ and proceed to higher orders,
\begin{equation}
\begin{aligned}
\rho &= \frac{1}{j} 
+ \epsilon \pare{
\frac{C_0 \, \cos (2 t)}{2 \sqrt{2}\, \pi \, j^2 }
- \sum_k \frac{c_k \, (1+k) \, \sin ((k+2) t) \, \Phi_{k,0,0} (\eta)}{j^2 (k+2)}
}
+ \sum_{n=2} \epsilon^n \, \rho_n (t, \eta)
\\[1mm]
\phi &= t 
+ \epsilon \pare{
\frac{C_0 \, \sin (2 t)}{4 \sqrt{2}\, \pi \, (1-j)}
- \sum_k \frac{c_k \, (1+k) \, \cos ((k+2) t) \, \Phi_{k,0,0} (\eta)}{2 (1-j) (k+2)}
}
+ \sum_{n=2} \epsilon^n \, \phi_n (t, \eta) .
\end{aligned}
\label{gen ansatz epsilon1}
\end{equation}

\bigskip
When $\partial_\eta \rho = \partial_\eta \phi = 0$, or equivalently if $c_k = 0$ for $k>0$, we could solve EoM at higher orders of $\epsilon$.
In this case, $\phi(t)$ is fixed by the angular momentum \eqref{conserved charges},
\begin{equation}
j = \frac{1}{\rho ^2}
+\frac{(\rho -1) R \dot \phi }{\rho ^{5/2} } \(
\kappa ^2-\frac{R^2 \dot \rho^2}{4 (\rho -1) \rho ^2}-\frac{(\rho -1) R^2 \dot \phi^2}{\rho } \)^{-1/2} \,.
\label{gen Jphi}
\end{equation}
We can solve the EoM for $\rho(t)$ as
\begin{align}
\rho &= \frac{1}{j}
+\epsilon \pare{ {\tt c} \cos (2 t)+{\tt d} \sin  (2 t) }
+\epsilon ^2 \, \frac{j (2 j-3) ({\tt c}^2+{\tt d}^2)}{2  (j-1)}
\notag \\[1mm]
&\quad -\epsilon ^3 \, \frac{3 j^2 ({\tt c}^2+{\tt d}^2) }{4 (j-1)} \pare{
({\tt d} +4 {\tt c} \, t) \, \sin (2 t) 
+ ({\tt c} -4  {\tt d} \, t) \, \cos (2 t) }
\notag \\[1mm]
&\quad -\epsilon ^4 \, \frac{j^3 ({\tt c}^2+{\tt d}^2) }{8  (j-1)^3} \pare{
(2 j (j (4 j-9)+6)-3)  ({\tt c}^2+{\tt d}^2)
+2 (j-1)^2 (({\tt d}^2-{\tt c}^2) \cos (4 t)
-2 {\tt c} {\tt d} \sin (4 t)) }
\notag \\[1mm]
&\quad + O(\epsilon^5) .
\label{rho sol 4th}
\end{align}
The parameters $({\tt c}, {\tt d})$ are arbitrary constants of $O(1)$ which may depend on $\epsilon$. 
The energy of an oscillating D3 brane is
\begin{equation}
E = \frac{N_c}{g_s} \, \pare{ j
+ \epsilon ^2 \, \frac{ ({\tt c}^2+{\tt d}^2) j^4 }{2 (1-j) }
- \epsilon ^4 \, \frac{ ({\tt c}^2+{\tt d}^2)^2 j^6 \( 7 (j-3) j+15 \) }{8 (1-j)^3 }
+ O(\epsilon^6) } .
\label{ED3 osc}
\end{equation}
The solution \eqref{rho sol 4th} contains secular terms like
\begin{equation}
(\text{polynomial of $t$}) \ \times \ \cos (2 m t), \quad
(\text{polynomial of $t$}) \ \times \ \sin (2 m t), \qquad (m \in \bb{Z}).
\end{equation}
We should renormalize the frequencies in order to keep $\rho(t)$ finite at large $t$.

\bigskip
When the first-order solution has KK modes, namely if $c_k \ne 0$, we do not find higher-order classical solutions.
One possible interpretation is that the solutions with non-trivial KK modes on S$^3$ are not purely classical, and hence they cannot produce $E- J = O(N_c) > 0$.
If we think of $\epsilon \gtrsim \lambda /N_c \sim g_s$\,, then the higher-order corrections are mixed up with $g_s$ corrections.

\section{Comments on AdS/CFT}\label{sec:AdS/CFT}

We look for the strong coupling counterpart of the all-loop ansatz \eqref{DG cont all-loop}
\begin{equation}
\Delta - J = \frac{\tilde f (\lambda)}{N_c} \sum_{\alpha=1}^{p-1} m_\alpha \,  \lambda_\alpha ( \{ n_{ij} \} )
\label{compare all-loop}
\end{equation}
where $\alpha$ labels the eigenvalues of the $(p-1)$ coupled oscillators, and $n_{ij} (\sigma)$ is a non-negative integer satisfying
\begin{equation}
n_{ij} (\sigma) = n_{i \to j} (\sigma) + n_{j \to i} (\sigma), \qquad
\sum_{j=1}^p n_{i \to j} = \sum_{j=1}^p n_{j \to i} \,, \qquad
n_Y = \sum_{i=1}^p \sum_{j=1}^p n_{i \to j} (\sigma).
\label{nij constraints_rev}
\end{equation}
The second equation is the Gauss law constraints \eqref{Gauss-law nij}, which suggests that $n_{ij} (\sigma)$ is the number of open strings stretching between the $i$-th and $j$-th branes.

At $p=2$, the ansatz \eqref{compare all-loop} becomes
\begin{equation}
\Delta - J = \frac{\tilde f (\lambda)}{N_c} \, m \, n_{12} \,, \qquad
n_{12} (\sigma) \in \bb{Z}_{\ge 0} \,, \qquad
m = 1,2, \dots, \left\lceil N_c - \frac{n_Z}{2} + 1 \right\rceil .
\label{compare all-loop, p=2}
\end{equation}

\subsection{Comparison with the oscillating D3-brane}\label{sec:oscillating}

In Section \ref{sec:strong coupling}, we found that there are two types of classical D-brane motion around the BPS configuration, whose energies are given by
\begin{equation}
E - J = 
\begin{cases}
\ds \frac{N_c}{g_s} \[
\epsilon ^2 \, \frac{ ({\tt c}^2+{\tt d}^2) j^4 }{2 (1-j) }
- \epsilon ^4 \, \frac{ ({\tt c}^2+{\tt d}^2)^2 j^6 \( 7 (j-3) j+15 \) }{8 (1-j)^3 }
+ O(\epsilon^6) \]
&\quad (k=0)
\\[5mm]
\ds \frac{N_c}{g_s} \[ 
\frac{\epsilon ^2 \, c_k^2 \, (k+1)^2}{8 (1-j) (k+2)} 
+ O(\epsilon^4)
\]
&\quad (k \ge 2) .
\end{cases}
\label{EJ kkgg}
\end{equation}
The first solution can be easily extended to higher orders of $\epsilon$, whereas the second solution cannot be extended to the next order by means of the simple separation of variables.

We argue that the oscillating D-brane should correspond to the harmonic oscillator of the effective $U(p)$ theory. More explicitly, we relate the energy of oscillating D-brane \eqref{EJ kkgg} at large $k$ and the all-loop ansatz \eqref{compare all-loop, p=2} at $p=2$ and large $m$,
\begin{equation}
E - J \simeq \frac{ N_c^2 \, \epsilon^2}{\lambda } \, \frac{ \pi \, c_k^2}{2(1-j)} \, k
\qquad \leftrightarrow \qquad
\Delta - J = \frac{\tilde f (\lambda)}{N_c} \, n_{12} (\sigma) \, m \,.
\label{conjecture ads/cft}
\end{equation}
where we used $\lambda \equiv R^4/\alpha'{}^2 = 4 \pi g_s N_c$\,.
We regard $\epsilon$ as the quantity of $O(1/N_c)$, which is an effect of the fundamental strings moving around the D3-brane.
 Then $k$ should be less than $O(\epsilon^{-1/2})$ to keep the corrections to $(E-J)$ small. 
This bound corresponds to the fact that the mode number $m$ is bounded from above at $O(N_c)$.

\bigskip
Let us present several lines of reasoning behind this identification.

Firstly, both dispersion relations are gapless, and one can excite the BPS state by supplying an arbitrarily small amount of energy.

Secondly, let us recall the AdS/CFT correspondence for the half-BPS states.
At weak coupling, the Young diagrams with different shapes start mixing at one-loop. The column length of a Young diagram can be interpreted as the radial direction of droplet patterns in the LLM plane \cite{Lin:2004nb}. This interpretation shows that the D-brane itself should oscillate.

Thirdly, we cannot deform the $j=1$ solution, i.e. the maximal giant graviton.
This corresponds to the fact that one cannot attach a box representing $Y$ to the Gauss graph operator $\cO^{R, r} (\sigma)$ if $r$ has the column of length equal to $N_c$ in \eqref{def:r1R}.

\begin{figure}[t]
\begin{equation*}
\begin{tikzpicture}[baseline={([yshift=-10mm] current bounding box.north)},x=10mm,y=10mm]
\draw [thick, dashed, ->] ({4*cos(240)},{4*sin(240)}) arc [start angle=240, end angle=300, radius=40mm];
\draw [thick, dashed, ->] ({5*cos(240)},{5*sin(240)}) arc [start angle=240, end angle=300, radius=50mm];
\coordinate (A) at (0,-4);
\coordinate (B) at (0,-5);
\draw[thick,decorate,decoration={coil,aspect=0}] (A) -- (B);
\shade[ball color=black] (A) circle (1.5mm);
\shade[ball color=white] (B) circle (1.5mm);
\end{tikzpicture}
\qquad \Longrightarrow \qquad
\begin{tikzpicture}[baseline={([yshift=-10mm] current bounding box.north)},x=10mm,y=10mm]
\draw [thick, dashed, ->] ({4*cos(240)},{4*sin(240)}) arc [start angle=240, end angle=300, radius=40mm];
\draw [thick, dashed, ->] ({5*cos(240)},{5*sin(240)}) arc [start angle=240, end angle=300, radius=50mm];
\shade[ball color=black] ({4*cos(250)},{4*sin(250)}) circle (1.5mm);
\shade[ball color=black] (0,-4) circle (4.9mm);
\shade[ball color=black] ({4*cos(290)},{4*sin(290)}) circle (1.5mm);
\shade[ball color=white] ({5*cos(250)},{5*sin(250)}) circle (1.5mm);
\shade[ball color=white] (0,-5) circle (4.9mm);
\shade[ball color=white] ({5*cos(290)},{5*sin(290)}) circle (1.5mm);
\end{tikzpicture}
\end{equation*}

\caption{(Left) Open string stretching between two D3-branes as a probe. (Right) D3-branes start expanding and shrinking.}
\label{fig:probe}
\end{figure}
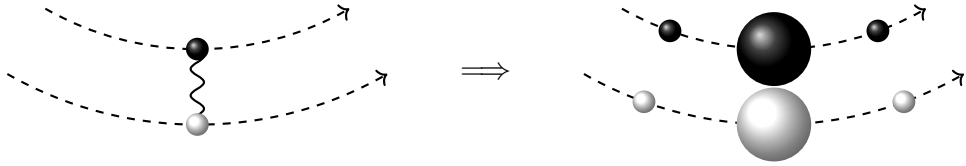

\bigskip
We make some comments on the parameter $n_{i \to j} (\sigma)$ in \eqref{nij constraints_rev}.

On the weak coupling side, the parameter $n_{i \to j} (\sigma)$ is interpreted as the number of open strings from the $i$-th brane to the $j$-th brane. 
On the strong coupling side, it is not clear whether we can introduce an open string as a probe, because non-maximal giants start oscillating by perturbing with infinitesimal energy; see Figure \ref{fig:probe}.
Moreover, the length of the probe string must be negligibly small, in order to maintain the gapless property of the dispersion relation. We will discuss a related issue in Section \ref{sec:reflecting}.

The parameter $n_{i \to j} (\sigma)$ also counts the number of $Y$-fields, which should correspond to (part of) the angular momentum in S$^3$ denoted by $J_Y = n_Y$\,.
At strong coupling, the D-branes wrapping S$^3$ inside S$^5$ have the zero angular momentum in $\theta_1$ due to the static gauge \eqref{sec:static gauge}.
This conclusion is reasonable because we do not see $J_Y \sim O(1)$ at strong coupling, in agreement with the assumption $n_Y \ll n_Z \sim O(N_c)$ in the all-loop ansatz. 
If we still want to explain $J_Y$\,, we may also add a point particle rotating S$^3$ carrying $J_Y$\,. This particle does not interact with D-branes at the leading order of large $N_c$\,.\footnote{If we think of the point particle as a closed string, this may also correspond to the term $n_{ii} (\sigma)$ in the effective $U(p)$ theory Hamiltonian, which shows up in the subleading order of large $N_c$ \cite{Koch:2013yaa,deCarvalho:2018xwx}.}

When $p>2$ as in \eqref{compare all-loop}, we should find $p$ D3-branes oscillating individually, corresponding to the $U(1)^p$ symmetry of the effective $U(p)$ theory.
The symmetry can be enhanced to non-abelian, e.g. $U(2) \times U(1)^{p-2}$, if some D-branes stay on top of each other at strong coupling, or if we give up the distant corners approximation at weak coupling. 
This point will be discussed in Appendix \ref{app:non-abelian}.

\subsection{On reflecting magnons}\label{sec:reflecting}

At strong coupling, there is another known situation of open strings ending on a giant graviton, called the reflecting magnons in the literature \cite{Hofman:2007xp}.
An example can be depicted as
\begin{equation}
\Biggl( \ \parbox{60mm}{Spinning open strings ending on the $Z=0$ maximal giant graviton} \ \Biggr) 
\quad = \quad
\begin{tikzpicture}[baseline={([yshift=-20mm]current bounding box.north)},x=1mm,y=1mm]
\coordinate (O) at (0,0);
\coordinate (A) at ({20*cos(210)},{20*sin(210)});
\coordinate (B) at ({20*cos(130)},{20*sin(130)});
\coordinate (C) at ({20*cos(30)},{20*sin(30)});
\coordinate (D) at ({20*cos(320)},{20*sin(320)});
\draw [thick, dashed, color=red] (O) -- (A);
\draw [thick, color=blue] (A) -- (B);
\draw [thick, color=blue] (B) -- (C);
\draw [thick, color=blue] (C) -- (D);
\draw [thick, dashed, color=red] (D) -- (O);
\draw [thick] (O) circle [radius=20];
\fill (O) circle [radius=1,fill=black];
\end{tikzpicture}
\quad
\label{fig:reflecting mag}
\end{equation}
where the red dashed lines represent boundary magnons, and the blue thick lines represent bulk magnons.
This figure is the same as the top view of Figure \ref{fig:oscillating giants 0} if the angular momentum $J_\phi$ takes the maximal value at $r=0$.
The energy of open strings in \eqref{fig:reflecting mag} having a large angular momentum $J_{\rm string}$ is given by the energy of an integrable open spin chain with $\alg{su}(2|2)$ symmetry \cite{Beisert:2005tm} as
\begin{equation}
E - J_{\rm string} = \sum_{\alpha=L,R} \sqrt{Q_\alpha^2 + \frac{\lambda}{\pi^2}}
+ \sum_{i=1}^M \sqrt{1 + \frac{\lambda}{\pi^2} \sin^2 \frac{p_i}{2} } \,,
\qquad (N_c \gg J_{\rm string} \gg 1)
\label{dispersion rmag}
\end{equation}
where $M$ is the number of bulk magnons.
Note that the boundary terms can be interpreted as extra magnons with $p = \pi$.

At strong coupling $\lambda \gg 1$, the magnon energy \eqref{dispersion rmag} is equal to the sum of the open string length multiplied by the string tension. This dispersion relation cannot be gapless, because an open string should connect the equator and the north pole of S$^5$.

At weak coupling, the system \eqref{fig:reflecting mag} is expected to be dual to a long operator attached to the determinant of $Z$'s,
\begin{equation}
\cO_{\rm det} = \sum_{\substack{i_1, i_2 , \dots, i_{N_c} , \\ j_1, j_2, \dots, j_{N_c}=1}}^{N_c}
\epsilon^{i_1i_2 \dots i_{N_c}}_{j_1 j_2 \dots j_{N_c}} \,
Z^{j_1}_{i_1} Z^{j_2}_{i_2} \dots Z^{j_{N_c-1}}_{i_{N_c-1}} \ 
\( {\color{red}{\chi_L}} \dots ZZ \dots {\color{blue}{\psi_1} }
\dots {\color{blue}{\psi_2}} \dots ZZ \dots {\color{red}{\chi_R}} \)^{j_{N_c}}_{i_{N_c}}
\label{def:Odet}
\end{equation}
where $\chi_L \,, \chi_R$ represent the boundary magnons and $\psi_1 \,, \psi_2 \,, \dots$ represent the bulk magnon.

Consider the expansion of the determinant-like operator \eqref{def:Odet} in the Gauss graph basis.
It is known that the determinant of $Z$ corresponds to $r = \spar{N_c}$, a single column of length $N_c$\,, and a single-trace operator is a linear combination of single hook Young diagrams \cite{Kristjansen:2002bb}.\footnote{This observation is true for the Schur polynomials, or the operators made out of $Z$'s only. The single-trace operators with magnons may correspond to the restricted Schur polynomials whose $R$ is almost (but not exactly) a single-hook.}
Thus we expect that the determinant-like operator \eqref{def:Odet} should be expanded by $O^{R,r} (\sigma)$, where both $R$ and $r$ consist of a single hook attached to the column of length $O(N_c)$.
We can generalize this system by introducing multiple giant gravitons.
The Young diagrams $(R, r)$ for Gauss graph basis \eqref{def:r1R} should be modified as
\begin{equation}
r_1 = \ 
\begin{tikzpicture}[baseline={([yshift=-10mm]current bounding box.north)},x=4mm,y=4mm]
\tikzmath{\x1=12; \x2=10; \x3=7.5; \x4=5.5; \x5=3; \y5=4;}
\draw (1,0) rectangle (2,-2*\x2);
\draw (2,0) rectangle (3,-2*\x3);
\draw (3,0) rectangle (4,-2*\x4);
\draw (4,0) -- (4+\y5,0);
\draw (4+\y5,0) -- (4+\y5,-1);
\draw (4+\y5,-1) -- (5,-1);
\draw (5,-1) -- (5,-2*\x5);
\draw (5,-2*\x5) -- (4,-2*\x5);
\node at (1.5,-\x2) {\small $l_1$};
\node at (2.5,-\x3) {\small $\vdots$};
\node at (3.5,-\x4) {\small $l_p$};
\end{tikzpicture} \ , \qquad
R =
\begin{tikzpicture}[baseline={([yshift=-10mm]current bounding box.north)},x=4mm,y=4mm]
\tikzmath{\x1=12; \x2=10; \x3=7.5; \x4=5.5; \x5=3; \y5=4;}
\draw (1,0) rectangle (2,-2*\x2);
\draw (2,0) rectangle (3,-2*\x3);
\draw (3,0) rectangle (4,-2*\x4);
\draw (4,0) -- (4+\y5,0);
\draw (4+\y5,0) -- (4+\y5,-1);
\draw (4+\y5,-1) -- (5,-1);
\draw (5,-1) -- (5,-2*\x5);
\draw (5,-2*\x5) -- (4,-2*\x5);
\draw [fill=gray!50] (1,-2*\x2) rectangle (2,-2*\x2-1);
\draw [fill=gray!50] (1,-2*\x2-1) rectangle (2,-2*\x2-2);
\draw [fill=gray!50] (2,-2*\x3) rectangle (3,-2*\x3-1);
\draw [fill=gray!50] (3,-2*\x4) rectangle (4,-2*\x4-1);
\draw [fill=gray!50] (4,-2*\x5) rectangle (5,-2*\x5-1);
\draw [fill=gray!50] (4+\y5,0) rectangle (5+\y5,-1);
\node at (1.5,-2*\x2-3) {\small $1$};
\node at (2.5,-2*\x3-2) {\small $\vdots$};
\node at (3.5,-2*\x4-2) {\small $p$};
\end{tikzpicture}
\label{def:r1R-rm}
\end{equation}
Recall that in the distant corners approximation, we can neglect the mixing of $Y$ fields in the different columns.
Thus, the mixing matrix of the system \eqref{def:r1R-rm} at large $N_c$ should factorize between the single-trace part and the effective $U(p)$ theory part,
\begin{equation}
\Delta - n_Z = \Delta (\text{Reflecting magnons})
+ \Delta (\text{Oscillating giants}) .
\label{Delta splits}
\end{equation}
We can interpret the first term as \eqref{dispersion rmag} and the second term as \eqref{conjecture ads/cft} if $J_{\rm string}$ (the length of a single-hook) is large.

\bigskip
Below are a few remarks on the $\alg{su}(2|2)$ symmetry.

We expect that the giant graviton possesses the residual superconformal symmetry $\alg{psu}(2|2)^2$, based on the $\kappa$-symmetric formulation of the D3-brane action on \AdSxS\ \cite{Metsaev:1998hf}.
However, we do not find any reasons that this symmetry should be promoted to the centrally-extended $\alg{su}(2|2)$. In other words, it is likely that the oscillating D-brane solutions are intrinsically non-BPS, and not centrally-extended BPS.

In \cite{deCarvalho:2020pdp}, they constructed the $\alg{su}(2|2)$ generators of the effective $U(p)$ theory. 
They proposed the central extension by
 \begin{equation}
\{ (Q^\alpha{}_a)_i ,(Q^\beta{}_b)_j\} =\epsilon^{\alpha\beta}\epsilon_{ab} \, P_{ij}
\qquad
\{ (S^a{}_\alpha)_i ,(S^b{}_\beta)_j\} =\epsilon_{\alpha\beta}\epsilon^{ab} \, K_{ij}
\label{wrong QS}
\end{equation}
and interpreted $\Delta_{ij}^\pm$ as the centers
\begin{equation}
P_{ij} = \alpha (d_i^+ - d_j^+) , \qquad
K_{ij} = \beta (d_i^- - d_j^-), \qquad
\Delta - J \stackrel{?}{=} \frac12 \, \sum_{ij} \sqrt{1 + P_{ij} \, K_{ij} }
\label{su22 proposal}
\end{equation}
which correspond to the second term of \eqref{Delta splits}.
However, the proposal \eqref{su22 proposal} is inconsistent because LHS of \eqref{wrong QS} is symmetric under $(a,\alpha,i) \leftrightarrow (b,\beta,j)$ whereas the RHS is anti-symmetric.\footnote{In addition, \eqref{su22 proposal} lacks $n_{ij} (\sigma)$, and disagrees with the two-loop result if $\alpha \beta = O(1)$.}

An alternative approach to the question of the central extension is as follows.
Here we gave several expressions of the D-brane energy; the conjectured all-loop formula \eqref{compare all-loop}, $\Delta (\text{Oscillating giants})$ in \eqref{Delta splits}, and the boundary term in the spin chain energy \eqref{dispersion rmag}.
Only the last expression takes the square-root form inherited from the centrally extended $\alg{su}(2|2)$ symmetry.\footnote{The bulk and boundary terms are essentially the same in the collective coordinate approach \cite{Berenstein:2013md,Berenstein:2013eya,Berenstein:2014zxa}.}
For the other cases we do not find square-roots, and one possible reason is the wrapping corrections to the $Z=0$ brane \cite{Correa:2009mz}.
Roughly speaking, the oscillating giants without open strings would correspond to the $J_{\rm string} \to 0$ limit of the integrable spin chain.
In other words, it may be possible to refine the proposal \eqref{su22 proposal} by studying the states like \eqref{def:r1R-rm} in the limit of $J_{\rm string} \gg 1$.\footnote{We thank an anonymous referee of JHEP for this comment.}

\section{Summary}

In this paper, we studied a non-planar large $N_c$ limit of $\cN=4$ SYM as a new example of the AdS/CFT correspondence.
First, we reviewed the Hamiltonian of an effective $U(p)$ theory coming from the perturbative dilatation operator acting on the Gauss graph basis. When $p=2$, this model is related to the finite harmonic oscillator.
Second, we proposed an all-loop ansatz based on the effective $U(p)$ theory. We found mutually commuting charges generated by the difference operators.
By taking the continuum limit, we argue that higher loop terms should be proportional to the one-loop result.

In our all-loop ansatz, the harmonic oscillators remain non-vanishing in the large $N_c$ limit, giving a gapless dispersion relation. In particular, it indicates that non-BPS excited giant gravitons should be continuously connected to the BPS giant graviton at strong coupling.

We investigated the classical D3-brane action on \AdSxS\ and found that a non-maximal spherical giant graviton can be excited in a gapless way. 
We argued that this new oscillating brane with KK modes on S$^3$ is a good candidate for the AdS/CFT dictionary which corresponds to the harmonic oscillator in the effective $U(p)$ theory.

\bigskip
Possible future directions are sketched as follows.

One direction is to investigate this correspondence further.
At weak coupling, the mixing matrix on the Gauss graph basis should be evaluated in a more general setup. This includes higher loop effects, a larger set of operators including the $\alg{sl}(2)$ sector \cite{deMelloKoch:2011vn}, and the corrections from higher orders in $n_Y/n_Z$ \cite{Ali:2015yrs}.
At strong coupling, the dynamics of D3-brane on \AdSxS\ should be studied in a comprehensive way. This includes to resum the $\epsilon$ series in the $k=0$ solution \eqref{rho sol 4th}, and to investigate the non-abelian DBI action \cite{Myers:1999ps}. The non-abelian analysis of the pp-wave background \cite{Sadri:2003mx}, the F1-D3 system \cite{Hashimoto:1997px} and the D1-D3 system \cite{Constable:1999ac} may be helpful. A closely related method is the matrix regularization of the 
worldvolume theory in the pp-wave background \cite{SheikhJabbari:2004ik,SheikhJabbari:2005mf,AliAkbari:2006gh}, which should capture part of the energy spectrum at strong coupling.

It is interesting to generalize the computation of the three-point function between two giants and one graviton, into the correlators with oscillating giants \cite{Bissi:2011dc,Lin:2012ey,Caputa:2012yj}, which serves as a non-trivial check of our proposal.

Another direction is to determine the general spectrum of the Hamiltonian of the effective $U(p)$ theory. This includes solving the finite oscillator for $p>2$, finding consistent wave-functions for AdS giants, and understanding the structure of $1/N_c$ corrections.

The role of the superconformal symmetry needs to be examined.
In particular, the $\cN=4$ SYM theory can be deformed while keeping $\alg{su}(2|2)^2$ \cite{Caetano:2020ofu}. 
It is worth investigating the corresponding deformation at strong coupling and finding the relation to the system of giant gravitons in \AdSxS.

A challenging question is whether the ``non-planar integrability'' can be found at strong coupling.
One starting point is the $\kappa$-symmetric D3-brane action in \AdSxS\ \cite{Metsaev:1998hf}. Then, through the reduction to two-dimensions \eqref{D3 action 2d}, we may be able to find a classical integrable system. The two-dimensional reduction may not be necessary if one can construct an integrable system in 4d along the line of \cite{Costello:2017dso,Costello:2018gyb,Costello:2019tri}.

\subsubsection*{Acknowledgments}

RS thanks the organizers of the workshops YITP-W-20-03 on {\it Strings and Fields 2020}, and {\it Online 2020 NTU-Kyoto high energy physics workshop} for stimulating discussions.
He is grateful to Robert de Mello Koch, Arkady Tseytlin and Keisuke Okamura for comments on the manuscript.
This research is supported by NSFC grant no. 12050410255.

\appendix

\section{Review of the Gauss graph basis}\label{app:Gauss}

We briefly review the construction of the Gauss graph basis, and how it simplifies the action of the perturbative one-loop dilatation operators of $\cN=4$ SYM.

\subsection{Notation}\label{app:notation}

Let $S_L$ be a permutation group of degree $L$. 
Its irreducible representations are labeled by a partition (Young diagram) $R \vdash L$, whose dimensions are denoted by $d_R$\,.
$D^R_{IJ} (\sigma)$ denotes the matrix representation of $\sigma$ in the irreducible representation $R$ with the component $(I,J)$, where $I, J=1,2, \dots, d_R$\,.

Consider the restriction $S_L \downarrow (S_m \otimes S_n)$ with $m+n=L$.
We denote the irreducible decomposition by
\begin{equation}
R = \bigoplus_{\substack{r \vdash m\\ s \vdash n}} \bigoplus_{\nu=1}^{g(r,s;R)} \( r \otimes s \)_\nu
\label{LR coeff by decomposition}
\end{equation}
where $\nu$ is a multiplicity label and $g(r,s;R)$ is the Littlewood-Richardson coefficient.
The branching coefficients are defined by the overlap between the components
\begin{equation}
\brT^{R \to (r, s), \nu}_{I \to (i, j)} = \Vev{ \atop{R}{I} \, \Big| \, \matop{r&s}{i&j}{\nu} } ,\qquad
B^{R \to (r, s), \nu}_{I \to (i, j)} = \Vev{ \matop{r&s}{i&j}{\nu}  \, \Big| \, \atop{R}{I} } \,.
\label{app:branching coeffs}
\end{equation}
See \cite{Suzuki:2020oce} for the properties of these quantities.

We denote partitions of an integer, or Young diagrams, in two ways.
The symbol $y = \spar{l_1, l_2, \dots, l_p}$ means that the $i$-th column of the Young diagram $y$ has the length $l_i$\,.
The symbol $y = [ m_1, m_2, \dots, m_q]$ means that the $j$-th row of $y$ has the length $m_j$\,.
It follows that
\begin{equation}
y \vdash L= \sum_{i=1}^p l_i = \sum_{j=1}^q m_j \,, \qquad
l_1 \ge l_2 \ge \dots \ge l_p \,, \qquad
m_1 \ge m_2 \ge \dots \ge m_q \,.
\label{app:Young notation}
\end{equation}

\subsection{Distant corners approximation}\label{app:distant}

We introduce the collective index
\begin{equation}
(Z^{\otimes n})^{\vec \imath}_{\vec \jmath} = Z^{i_1}_{j_1} \, Z^{i_2}_{j_2} \, \dots \, Z^{i_n}_{j_n} \,, \qquad
(U_\alpha)^{\vec \jmath}_{\vec k} = \delta^{j_1}_{k_{\alpha(1)}} \, \delta^{j_2}_{k_{\alpha(2)}} \, \dots \, \delta^{j_n}_{k_{\alpha(n)}} \quad (\alpha \in S_n)
\label{app:collective ind}
\end{equation}
with $i_p, j_p = 1,2, \dots, N_c$\,. The matrix $U_\alpha$ satisfies the composition rules
\begin{equation}
(U_\alpha)^{\vec \imath}_{\vec \jmath} \, (U_\beta)^{\vec k}_{\vec \imath} 
= (U_{\alpha \beta})^{\vec k}_{\vec \jmath} \,, \qquad
(U_\alpha)^{\vec \imath}_{\vec \jmath} \, (U_\beta)^{\vec \jmath}_{\vec k} 
= (U_{\beta \alpha})^{\vec \imath}_{\vec k}
\\[1mm]
\trb{\, n} (U_{\alpha}) = N_c^{C(\alpha)} .
\label{app:trace SW dual}
\end{equation}
We denote multi-trace operators in the $\alg{su}(2)$ sector by
\begin{equation}
\trb{\,L} \( U_\alpha \cdot Y^{\otimes n_Y} Z^{\otimes n_Z}\)
= \sum_{ i_1, i_2, \dots, i_L =1}^{N_c} 
Y^{i_1}_{i_{\alpha(1)}} \, Y^{i_2}_{i_{\alpha(2)}} \, \dots \, Y^{i_{n_Y}}_{i_{\alpha(n_Y)}} 
Z^{i_{n_Y+1}}_{i_{\alpha(n_Y+1)}} \, Z^{i_{n_Y+2}}_{i_{\alpha(n_Y+2)}} \, \dots \, Z^{i_L}_{i_{\alpha(L)}} 
\end{equation}
with $L = n_Y + n_Z$ and $\alpha \in S_L$\,.
We define the restricted Schur basis of operators by
\begin{align}
\cO^{R,(r,s), \nu_+,\nu_-} &= \frac{1}{n_Y! n_Z!}
\sum_{\alpha \in S_L} \chi^{R,(r,s), \nu_+,\nu_-} (\alpha) \, 
\trb{\,L} \( U_\alpha \cdot Y^{\otimes n_Y} Z^{\otimes n_Z}\)
\label{app:restricted Schur su(2)} \\[1mm]
\chi^{R,(r,s), \nu_+,\nu_-} (\alpha) &=
\sum_{I,J=1}^{d_R} \sum_{i=1}^{d_r} \sum_{j=1}^{d_s} 
B^{R \to (r,s) \nu_+}_{I \to (i,j)} \, 
\brT^{R \to (r,s), \nu_-}_{J \to (i,j)} \,
D^R_{IJ} (\alpha) 
\label{app:restricted ch}
\end{align}
coming from the restriction $S_L \downarrow (S_{n_Y} \otimes S_{n_Z})$.

It is expected that the half-BPS operators dual to $p$ spherical giant gravitons consist of $p$ long columns, with $n_Z = O(N_c)$ with $N_c \gg 1$. Non-BPS operators can be constructed by attaching $Y$ fields.
We write $r = \spar{l_1, l_2 , \dots , l_p}$ where $l_i$ is the length of the $i$-th column.
In the distant corners approximation, we assume that $l_i - l_{i-1} \gg 1$, so that the corners of $r$ are well separated. Therefore, we typically work with Young diagrams
\begin{equation}
r = \ 
\begin{tikzpicture}[baseline={([yshift=-10mm]current bounding box.north)},x=4mm,y=4mm]
\tikzmath{\x1=12; \x2=9; \x3=6; \x4=3.5;}
\draw (0,0) rectangle (1,-2*\x1);
\draw (1,0) rectangle (2,-2*\x2);
\draw (2,0) rectangle (3,-2*\x3);
\draw (3,0) rectangle (4,-2*\x4);
\node at (0.5,-\x1) {\small $l_1$};
\node at (1.5,-\x2) {\small $l_2$};
\node at (2.5,-\x3) {\small $\vdots$};
\node at (3.5,-\x4) {\small $l_p$};
\end{tikzpicture} \ , \qquad
R =
\begin{tikzpicture}[baseline={([yshift=-10mm]current bounding box.north)},x=4mm,y=4mm]
\tikzmath{\x1=12; \x2=9; \x3=6; \x4=3.5;}
\draw (0,0) rectangle (1,-2*\x1);
\draw (1,0) rectangle (2,-2*\x2);
\draw (2,0) rectangle (3,-2*\x3);
\draw (3,0) rectangle (4,-2*\x4);
\draw [fill=gray!50] (0,-2*\x1) rectangle (1,-2*\x1-1);
\draw [fill=gray!50] (0,-2*\x1-1) rectangle (1,-2*\x1-2);
\draw [fill=gray!50] (1,-2*\x2) rectangle (2,-2*\x2-1);
\draw [fill=gray!50] (2,-2*\x3) rectangle (3,-2*\x3-1);
\draw [fill=gray!50] (3,-2*\x4) rectangle (4,-2*\x4-1);
\node at (0.5,-2*\x1-3) {\small $1$};
\node at (1.5,-2*\x2-2) {\small $2$};
\node at (2.5,-2*\x3-2) {\small $\vdots$};
\node at (3.5,-2*\x4-2) {\small $p$};
\end{tikzpicture}
\label{def:r1R}
\end{equation}
where we construct $R$ by adding $s \vdash n_Y$ (gray boxes) to $r \vdash n_Z$ (white boxes).

\subsection{Gauss graph basis}\label{sec:GGB}

We introduce the Gauss graph basis following \cite{deMelloKoch:2012ck}.

\subsubsection{Skew Young diagrams}

We can specify the representation of $Y$ fields in two ways, $s$ or $R / r$.
The states of $s$ are labeled by the standard Young tableaux, and those of $R /r$ are by the skew Young tableaux.
In the restricted Schur polynomial \eqref{app:restricted ch}, we may keep track of which box of $s$ goes to which box of $R / r$, before summing over the indices $(I,J,i,j)$.

In the distant corners approximation, $R / r$ consists of $p$ columns well separated from each other.
This indicates that only the column position, $1,2, \dots, p$, should be important in finding the eigenstates of the perturbative dilatation operator of $\cN=4$ SYM.

Consider an example of $p=3$, with $s= \spar{4,2,1}$ and $R / r = \spar{3,2,2}$.
We parameterize a state of $s$ and $R/r$ using only the column labels, as
\ytableausetup{centertableaux,boxsize=4.5mm}
\begin{align}
s = \quad \begin{ytableau}
*(gray!30) 1 & *(gray!30) 2 & *(gray!30) 3 \\
*(gray!30) 1 & *(gray!30) 2\\
*(gray!30) 1 \\
*(gray!30) 1
\end{ytableau}  \ , \qquad
R / r = \quad
\begin{ytableau}
\none & \none & *(gray!30) 1 \\
\none & \none & *(gray!30) 2 \\
\none & *(gray!30) 1 \\
\none & *(gray!30) 2 \\
*(gray!30) 1 \\
*(gray!30) 1 \\
*(gray!30) 3
\end{ytableau} 
\label{app:sRr example}
\end{align}
In other words, we project the standard Young diagrams of shape $s \vdash n_Y$ onto the trivial (totally symmetric) representation of
\begin{equation}
H = S_{s_1} \otimes S_{s_2} \otimes  \dots S_{s_p} \ \subset S_{n_Y} \,, \qquad
s = \spar{ s_1, s_2, \dots, s_p} \,.
\label{def:Hs}
\end{equation}

The group $H$ is an extra symmetry that emerges in the distant corners approximation \cite{deMelloKoch:2020agz}.
We should refine the label of the restricted Schur operator $\cO^{R,(r,s), \nu_+,\nu_-}$ by adding $\vec s = (s_1, s_2, \dots, s_p)$ which specifies how $s \vdash n_Y$ shows up in the skew Young diagram $R /r$.

\subsubsection{Adjacency matrix}

We can define the adjacency matrix $n_{i \to j}$ by counting how many $i$'s appear in the $j$-th column of the skew tableau $R/r$. In the above example \eqref{app:sRr example} we find
\begin{equation}
\{ n_{i \to j} \} = 
\begin{pmatrix}
2 & 1 & 1 \\
0 & 1 & 1 \\
1 & 0 & 0
\end{pmatrix} .
\end{equation}
The adjacency matrix satisfies a conservation law.
When a box with the label $i$ goes to the $j$-th column ($j \neq i$), then there must be a box $k$ which comes to the $i$-th column.
This implies the relation called the Gauss law constraints,
\begin{equation}
\sum_{j=1}^p n_{i \to j} = \sum_{j=1}^p n_{j \to i} \,.
\label{Gauss-law nij}
\end{equation}
Intuitively, we may regard the diagonal elements $n_{i \to i}$ as the number of closed strings on the $i$-th brane, and the off-diagonal elements $n_{i \to j}$ as the number of open strings between the $i$-th and $j$-th brane.

Here we defined the adjacency matrix from a skew Young diagram. The skew Young diagram is in one-to-one correspondence with the Gelfand-Testlin basis, as discussed in \cite{Koch:2011hb}.

\subsubsection{Permutation and the double coset}

We can determine the adjacency matrix $\{ n_{i \to j} \}$ from a permutation element $\sigma \in S_{n_Y}$ as follows.
Given $s = \spar{ s_1, s_2, \dots, s_p}$, we introduce a state
\begin{equation}
\begin{aligned}
\ket{ \vec s} &\equiv \ket{ \, 1 \, \bar \otimes \dots \bar \otimes \, 1 \, } \otimes 
\ket{ \, 2 \, \bar \otimes \dots \bar \otimes \, 2 \, } \otimes \dots\otimes
\ket{ \, p \, \bar \otimes \dots \bar \otimes \, p \, } 
\\[1mm]
&= \ket{ \, 
1^{\bar \otimes s_1} \otimes 
2^{\bar \otimes s_2} \otimes \dots \otimes 
p^{\bar \otimes s_p} \, } 
\end{aligned}
\end{equation}
where $\bar \otimes$ represents the symmetrized tensor product and $\otimes$ is the usual tensor product.
The symmetrization is equivalent to a sum over the states in $V_p^{\otimes n_Y}$ where $V_p = \{ 1,2, \dots, p \}$.
We can permute the state $\ket{ \vec s}$ by applying $\sigma \in S_{n_Y}$ to each summand as
\begin{align}
\ket{ \sigma \,, \vec s} 
&= \ket{ \sigma(1) \otimes \sigma(2) \otimes \dots \otimes \sigma(n_Y) }_{\rm symm}
\label{def:double coset s} \\[1mm]
&\equiv 
\ket{ \sigma(1) \, \bar\otimes \dots \bar\otimes \, \sigma(s_1) } \otimes 
\ket{ \sigma(s_1+1) \, \bar\otimes \dots \bar\otimes \, \sigma(s_1+s_2) } \otimes \dots\otimes
\ket{ \sigma(n_Y - s_p +1) \, \bar\otimes\dots \bar\otimes \, \sigma(s_Y) } .
\notag 
\end{align}
This result consists of $p$ tensor product of symmetrized components. It makes sense to count the number of $i$'s in the $j$-th symmetrized component, and call it $n_{i \to j}$\,.

Owing to the symmetrization, the action of the permutation $\sigma \in S_{n_Y}$ reduces to the action of an element in the double coset $\bar \sigma \in H \backslash S_{n_Y} / H$,
\begin{equation}
\bar \sigma = \frac{1}{|H|} \sum_{\gamma_1, \gamma_2 \in H} \gamma_1 \sigma \gamma_2 \,, \qquad
\sigma \in S_{n_Y} \,.
\end{equation}
We can compute $n_{i \to j}$ graphically as
\begin{equation}
\begin{tikzpicture}[baseline=(current bounding box.center),x=4mm,y=4mm]
\tikzmath{\x1=5; \x2=10; \x3=15; \x4=21; \y1=-1.5; \y2=-4.8; \y3=-8.1; \y4=-11.4;}
\draw (-1,0) -- (\x4,0);
\draw (-1,\y4) -- (\x4,\y4);
\foreach \ia in {1,2,3} {
\draw [{Latex[length=2mm]}-] (\ia,0) -- (\ia,\y4);
\node at (\ia, \y4-1) {\small $\ia$};
}
\foreach \ib in {6,8} \draw [{Latex[length=2mm]}-] (\ib,0) -- (\ib,\y4);
\node at (6, \y4-1) {\small $4$};
\node at (8, \y4-1) {\small $5$};
\foreach \ic in {16,19} \draw [{Latex[length=2mm]}-] (\ic,0) -- (\ic,\y4);
\node at (16, \y4-1) {\small $(n_Y-1)$};
\node at (19, \y4-1) {\small $n_Y$};
\foreach \ja in {0,5} {
\draw [fill=white] (\ja,\y1) rectangle (\ja+4,\y1-1.8);
\draw [fill=white] (\ja,\y3) rectangle (\ja+4,\y3-1.8);
}
\draw [fill=white] (15,\y1) rectangle (20,\y1-1.8);
\draw [fill=white] (15,\y3) rectangle (20,\y3-1.8);
\draw [fill=white] (0,\y2) rectangle (20,\y2-1.8);
\node at (9.5,\y2-0.8) {$\sigma$};
\node at (12.2,\y1-0.8) {$\dots$};
\node at (12.2,\y3-0.8) {$\dots$};
\node at (2.2,\y1-0.8) {\small $\gamma_2^{(1)}$};
\node at (7.2,\y1-0.8) {\small $\gamma_2^{(2)}$};
\node at (17.7,\y1-0.8) {\small $\gamma_2^{(p)}$};
\node at (2.2,\y3-0.8) {\small $\gamma_1^{(1)}$};
\node at (7.2,\y3-0.8) {\small $\gamma_1^{(2)}$};
\node at (17.7,\y3-0.8) {\small $\gamma_1^{(p)}$};
\node at (-6.5,\y2) {$\ds n_{i \to j} (\sigma) \sim \frac{1}{|H|^2} \sum_{\gamma_1 , \gamma_2 \in H}$};
\end{tikzpicture}
\label{fig:Nij-sigma}
\end{equation}
which corresponds to $(s_1, s_2, \dots, s_p) = (3,2, \dots, 2)$.

\subsubsection{Operator mixing in the Gauss graph basis}

We define operators in the Gauss graph basis by\footnote{Our definition looks slightly different from \cite{deMelloKoch:2012ck} because our restricted Schur basis \eqref{app:restricted Schur su(2)} is not normalized.}
\begin{equation}
O^{R,r}(\sigma) = |H| \sqrt{n_Y!} \sum_{j,k} \sum_{s \, \vdash n_Y} \sum_{\nu_-,\nu_+}
D^s_{jk}(\sigma ) \, B^{s\to 1_H, \nu_-}_{j} \, \brT^{s\to 1_H, \nu_+}_{k} 
\cO^{R,(r,s), \nu_+,\nu_-} .
\label{app:gauss graph basis}
\end{equation}

We consider the case $n_Z = O(N_c)$ and $n_Y = O(1)$, which should correspond to excited multiple giant gravitons.
The one-loop dilatation acting on the restricted Schur polynomial factorizes into the mixing of $Y$'s and the mixing of $Z$'s at the leading order of large $N_c$\,.\footnote{The factorization property is violated at the subleading order of $n_Y/n_Z$ \cite{Ali:2015yrs}.}
As shown in \cite{deMelloKoch:2012ck}, the mixing of $Y$'s can be solved by taking the Gauss graph basis. The eigenvalues are labeled by the symmetrized adjacency matrix,
\begin{equation}
n_{ij} (\sigma) \equiv n_{i \to j} (\sigma)  + n_{j \to i} (\sigma) , \qquad
n_{ij} (\sigma) = n_{ji} (\sigma).
\label{app:nij symmetric}
\end{equation}
The mixing of $Z$'s changes the shape of $r = \spar{l_1, l_2, \dots, l_p}$. We use the simplified notation
\begin{equation}
O(\vec l) = O(l_1, l_2, \dots, l_p), \qquad
\sum_{i=1}^p l_i = n_Z 
\label{app:Ovec l}
\end{equation}
in place of \eqref{app:gauss graph basis}.

\section{Explicit one-loop spectrum}\label{app:explicit spectrum}

We studying the spectrum of $D_1^G = -\cH_1$ following \cite{Carlson:2011hy,Koch:2011hb} with minor improvement.

We consider both continuum and discrete cases. 
The dilatation operator reduces to a set of harmonic oscillators with boundary conditions in the continuum limit \eqref{def:sqrt Nc limit}. In order to determine the dilatation spectrum before taking the limit, we solve the discrete case. The discrete case has been solved for $p=2$ by using the finite oscillator \cite{APW05,Jafarov:2011fu}. 
The spectra of the two cases agree, implying that the one-loop dimensions do not depend on the details of the continuum limit.

\subsection{Continuum case}\label{sec:cont case}

The spectrum of one-loop dilatation in the continuum limit for general $p$ has been studied in \cite{Koch:2011hb}.
In this limit, we find $D_1^G = - \cH_1 \to - \cD$ where
\begin{equation}
\cD \, F (\vec y) = E (\{ n_{ij} \}) \, F (\vec y), \qquad
\cD \equiv \sum_{i \neq j}^p n_{ij} (\sigma) \, \cD_{ij} \,.
\label{continuous Schroedinger}
\end{equation}
We solve this equation in the region of $\{ y_i \}$ given in \eqref{def:yi}. 
We rewrite the differential operator $\cD_{ij}$ in \eqref{def:cDij} as
\begin{equation}
\cD_{ij} \, F = \( A^+ (y_{ij}) \, A^- (y_{ij}) + \frac12 \) F, \qquad
A^\pm (y) = \frac{1}{\sqrt{\alpha}} 
\pare{ \frac{\alpha y}{2} \pm \frac{\partial}{\partial y} }
\label{def:Apm y}
\end{equation}
where $y_{ij} = y_i - y_j$\,. The new differential operators $\{ A^\pm (y) \}$ satisfy 
\begin{equation}
[ A^+ (y_{ij}) , A^- (y_{kl}) ] = \delta_{ik} + \delta_{jl} - \delta_{il} - \delta_{jk} \,.
\label{CCR Aijpm}
\end{equation}
The operator $\cD_{ij}$ has the symmetry
\begin{equation}
[ \cD_{ij} \,, \Lambda^\pm ] = 0, \qquad
\Lambda^+ \equiv \sum_{i=1}^p y_i \,, \qquad
\Lambda^- \equiv \sum_{i=1}^p \frac{\partial}{\partial y_j} \,.
\end{equation}
Thus, $\Lambda^\pm$ represent the zero modes of $\cD$.
Roughly speaking, $\Lambda^\pm$ correspond to the addition or removal of a box from each of the $p$ columns, suggesting that the spectrum of $\cH_1$ depends only on the difference of column lengths.

From \eqref{def:Apm y} one finds that the eigenvalues of $\cD_{ij}$ are written as $(m + \frac12)$.
The mode number $m$ should be chosen so that the variables $y_{ij}$ satisfy the Young diagram constraints \eqref{def:yi}.
These constraints are expressed in terms of the variables $( y_{12} \,, \dots \,, y_{p-1,p} \,, y_p )$ as
\begin{equation}
y_{12} \ge 0, \ \ \ y_{23} \ge 0, \ \ \ \dots \,, \ \ \ y_{p-1,p} \ge 0, \ \ \ y_p \ge 0.
\label{def:YD constraints}
\end{equation}
Consider the operator\footnote{We used $\sum_{i \neq j} n_{ij} (\sigma) = 2 n_Y$\,.}
\begin{equation}
\tilde \cD \equiv \cD - n_Y
= 2 \sum_{i < j}^p n_{ij} (\sigma) \, A^+ (y_{ij}) A^- (y_{ij}), \qquad
A^\pm (y_{ij}) = \sum_{k=i}^{j-1} A^\pm (y_{k,k+1}) .
\end{equation}
This operator can be written as the following quadratic form
\begin{equation}
\tilde \cD = 2 \sum_{u,v=1}^{p-1} \scr{M}_{u,v} \, A^+ (y_{u,u+1}) A^- (y_{v,v+1})
\end{equation}
where
\begin{equation}
\scr{M}_{u,v} = \begin{cases}
\ds \sum_{k=1}^v \sum_{l=u+1}^p n_{k,l} &\qquad (u \ge v) \\[5mm]
\ds \sum_{k=1}^u \sum_{l=v+1}^p n_{k,l} &\qquad (u < v) .
\end{cases}
\end{equation}
Note that $\scr{M}$ is a $(p-1) \times (p-1)$ matrix which depends on $\{ n_{i,j} \}$ with $i < j$.
By diagonalizing $\scr{M}$, we obtain the non-zero eigenvalues of $\tilde \cD$\,.
\begin{equation}
\tilde \lambda_u \, \delta_{uv} = \( \scr{S} \scr{M} \scr{S}^T \)_{uv} \,, \qquad
\tilde \cD = 2 \sum_{a=1}^{p-1} \tilde \lambda_a \, A^+ (\tilde z_a) A^- (\tilde z_a) .
\label{decoupled oscillator tcH}
\end{equation}
The new coordinates $\{ \tilde z_a \}$ are written as
\begin{equation}
y_{a,a+1} = \sum_{b=1}^p \, ( \scr{S}^T )_{ab} \, \tilde z_a \,.
\label{def:z to y}
\end{equation}

We define the vacuum of $\cH_1$ by requiring
\begin{equation}
A^- (\tilde z_a) \ket {0} = 0 \qquad (\forall a)
\end{equation}
and define excited states by applying the creation operators $A^+ (\tilde z_a)$.
We solve the Young diagram constraints \eqref{def:YD constraints} as follows.
The boundary of the constraints lies on $y_{a,a+1}=0$, or equivalently $\tilde z_a=0$. 
The wave function $\psi(\{ \tilde z_a\})$ should vanish at $\tilde z_a=0$ for all $a$, because it is ill-defined in the region $y_{a,a+1}<0$.
Recall that the creation operators are parity odd,
\begin{equation}
A^+ (- \tilde z_a)  = - A^+ (\tilde z_a) .
\end{equation}
The physical states should contain an odd number of creation operators for each $a$. Thus, the spectrum of $\tilde \cD$ is
\begin{equation}
\tilde \cD \, \psi_{\vec m} (\{ \tilde z_a\}) 
= 2 \sum_{a=1}^{p-1} \( 2m_a + 1\) \tilde \lambda_a \, \psi_{\vec m} (\{ \tilde z_a\}), \qquad
m_a \in \bb{Z}_{\ge 0} \,.
\end{equation}
The eigenvalues of the original equation \eqref{continuous Schroedinger} are
\begin{equation}
E (\{ n_{ij} \}) = n_Y + 2 \sum_{a=1}^{p-1} \( 2m_a + 1\) \tilde \lambda_a ( \{ n_{ij} \} ) .
\label{continuous Sch ev}
\end{equation}

\subsection{Discrete case}\label{sec:disc case}

The main difficulty in computing the discrete spectrum lies in how to impose the Young diagram constraints.
The Fock space created by the oscillator representation of $\cH_1$ in \eqref{def:cH1} is not useful, because it does not immediately solve the constraints.
Instead, we directly look for the wave functions.
Some functional identities are summarized in Appendix \ref{app:hypergeom}.

For simplicity, we consider the case of $p=2$. 
We take the linear combination of $O(l_1,l_2)$ in \eqref{def:Delta ij} as
\begin{equation}
\cO_f = \sum_{x=-l_2}^{\lceil (l_1-l_2)/2 \rceil} f(x) O(l_1-x , l_2+x) .
\end{equation}
The operator $\cH_{1,12}$ in \eqref{def:cH1} acts on $\cO_f$ as
\begin{multline}
\cH_{1,12} \, \cO_f = \Big( h(1, l_1) + h(2, l_2) \Big) \, \cO_f 
- \sqrt{ h(1, l_1) \, h(2, l_2+1)} \, \sum_{x} f(x) O( l_1-x-1, l_2+x+1)
\\
- \sqrt{ h(1, l_1+1) \, h(2, l_2)} \, \sum_{x} f(x) O( l_1-x+1, l_2+x-1) 
\end{multline}
which gives the following discrete eigenvalue equation
\begin{equation}
\begin{gathered}
\cH_{1,12} = h(1, l_1) + h(2, l_2) 
- \sqrt{ h(1, l_1) \, h(2, l_2+1)} \, e^{- \partial_x}
- \sqrt{ h(1, l_1+1) \, h(2, l_2)} \, e^{+ \partial_x} 
\\[2mm]
\cH_{1,12} \, f(x) = E \, f(x) .
\end{gathered}
\label{p=2 disc eq}
\end{equation}
This equation determines the eigenvalue of $D_1^G = -\cH_1$ at $p=2$ as
\begin{equation}
D_1^G = - 2 n_{12} \, E 
\label{discrete eigenvalue DE}
\end{equation}
where we used $n_{12} = n_{21}$\,.

We solve \eqref{p=2 disc eq} by relating it to a finite oscillator \cite{APW05,Jafarov:2011fu}.
Let us take a basis of states in the irreducible representations of $\alg{su}(2)$,
\begin{equation}
J_3 \, \ket{ j, j_3 } = j_3 \, \ket{ j, j_3 } , \qquad (j_3 = -j, -j+1, \dots, j)
\end{equation}
and a rotated basis
\begin{equation}
J_1 \, \ket{j, j_1}_1 = j_1 \, \ket {j, j_1}_1 \,, \qquad (j_1 = -j, -j+1, \dots, j).
\end{equation}
In view of the effective $U(p)$ theory, the $\alg{su}(2)$ generators can be interpreted as
\begin{equation}
J_1 = \frac12 \( d_1^+ \, d_2^- + d_2^+ \, d_1^- \), \qquad
J_2 = \frac{i}{2} \( d_1^+ \, d_2^- - d_2^+ \, d_1^- \), \qquad
J_3 = \frac12 \( d_1^+ \, d_1^- - d_2^+ \, d_2^- \) .
\end{equation}
The generator $(2 J_1)$ is identical to $\cJ_{12}$ in \eqref{def:cJij}.
The rotation matrix is given by
\begin{equation}
{}_1 \vev{ j , j_1 \, \big| \, j, j_3} 
= \frac{(-1)^{j+j_3}}{2^j} \, \sqrt{ \binom{2j}{j+j_3} \binom{2j}{j+j_1} } \ 
{}_2 F_1 (-j-j_3, -j-j_1 ; -2j; 2).
\label{def:su2 rotation mat}
\end{equation}
The $\alg{su}(2)$ generators acts on the states $\ket{j, j_1}_1$ in the standard way,
\begin{equation}
J_1 \ket{j, j_1}_1 = j_1 \ket {j,j_1}_1 \,, \qquad
J_\pm \ket {j, j_1}_1 = \sqrt{ (j \mp j_1) (j \pm j_1 + 1) } \ket{j , j_1 \pm 1}_1 \,.
\label{su2 oscillators on rotated basis}
\end{equation}

We define
\begin{equation}
n_Z \equiv l_1 + l_2 \,, \qquad
l_{12} = l_1 - l_2
\end{equation}
and assume $n_Z \le 2 N_c$\,, which is trivial at $p=2$.
Let us assign
\begin{equation}
j = N_c - \frac{n_Z - 1}{2} \,, \qquad
j_1 = \frac{- l_{12} - 1}{2} \,,\qquad
j + j_3 = m - 1
\end{equation}
where the new variables run the ranges
\begin{equation}
n_Z - 2 N_c - 2 \le l_{12} \le 2 N_c - n_Z \,, \qquad
1 \le m \le 2 N_c - n_Z + 2.
\end{equation}
Note that
\begin{equation}
l_{12} = n_Z - 2 N_c - 2 \quad \Leftrightarrow \quad l_2 = N_c + 1, \qquad
l_{12} = 2 N_c - n_Z \quad \Leftrightarrow \quad l_1 = N_c \,.
\end{equation}
The equations \eqref{su2 oscillators on rotated basis} become
\begin{equation}
\begin{aligned}
J_1 \, \Big| N_c - \frac{n_Z - 1}{2} \,, \frac{- l_{12} - 1}{2} \, \Big\rangle_1 &=  \frac{- l_{12} - 1}{2} \
\Big| N_c - \frac{n_Z - 1}{2} \,, \frac{- l_{12} - 1}{2} \, \Big\rangle_1
\\[1mm]
J_+ \, \Big| N_c - \frac{n_Z - 1}{2} \,, \frac{- l_{12} - 1}{2} \, \Big\rangle_1 &=  \sqrt{ ( N_c - l_2 + 1 ) (N_c-l_1 + 1) } \ 
\Big| N_c - \frac{n_Z - 1}{2} \,, \frac{- l_{12} + 1}{2} \, \Big\rangle_1
\\[1mm]
J_- \, \Big| N_c - \frac{n_Z - 1}{2} \,, \frac{- l_{12} - 1}{2} \, \Big\rangle_1 &= \sqrt{ (N_c - l_1) ( N_c - l_2 + 2 )  } \ 
\Big| N_c - \frac{n_Z - 1}{2} \,, \frac{- l_{12} - 3}{2} \, \Big\rangle_1 \,.
\end{aligned}
\label{su2 oscillators on rotated basis2}
\end{equation}
The last two lines agree with the off-diagonal terms in \eqref{p=2 disc eq} with the help of \eqref{def:hili}.

We can immediately solve \eqref{p=2 disc eq} by relating the wave function to the rotation matrix \eqref{def:su2 rotation mat}.
If we define
\begin{multline}
F_m (l_1,l_2) =
(-1)^{m-1} \, 2^{-N_c + \frac{1}{2} (n_Z-1)} 
\sqrt{\binom{2 N_c-n_Z+1}{m-1} \binom{2 N_c-n_Z+1}{N_c-l_1}} \ \times
\\[1mm]
{}_2F_1\left( -m+1, -N_c+l_1 ; -2 N_c + n_Z -1 ; 2 \right) .
\label{def:Fml}
\end{multline}
this function satisfies the recursion relation
\begin{multline}
\sqrt{ h(1, l_1) \, h(2, l_2+1)} \, F_m(l_1-1,l_2+1)
+ \sqrt{ h(1, l_1+1) \, h(2, l_2)} \, F_m (l_1+1,l_2-1)
\\
- \Big( h(1, l_1) + h(2, l_2) \Big) F_m (l_1,l_2)
= - 2 m \, F_m (l_1,l_2) 
\label{discrete eqs for Fm}
\end{multline}
which is equivalent to the discrete eigenvalue equation \eqref{p=2 disc eq} with $E=2m$.

We should impose the Young diagram constraints at $p=2$, namely
\begin{equation}
\begin{cases}
N_c \ge l_1 \ge l_2 \ge (N_c - l_1) &\qquad (n_Z \ge N_c)
\\[1mm]
 l_1 \ge l_2 \ge 0 &\qquad (n_Z \le N_c).
\end{cases}
\label{refined p=2 YD constraints}
\end{equation}
Note that the operator mixing does not change the value of $n_Z = l_1 + l_2$\,.
It turns out that our solution \eqref{def:Fml} can solve these constraints only in limited cases.

\begin{figure}[t]
\begin{center}
\includegraphics[scale=0.7]{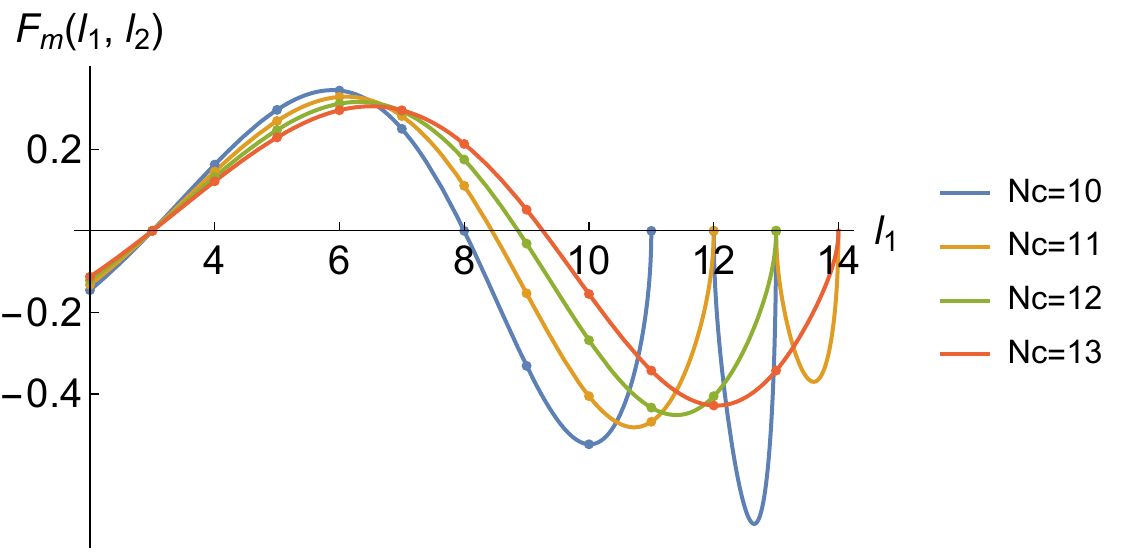}
\hspace{7mm}
\includegraphics[scale=0.7]{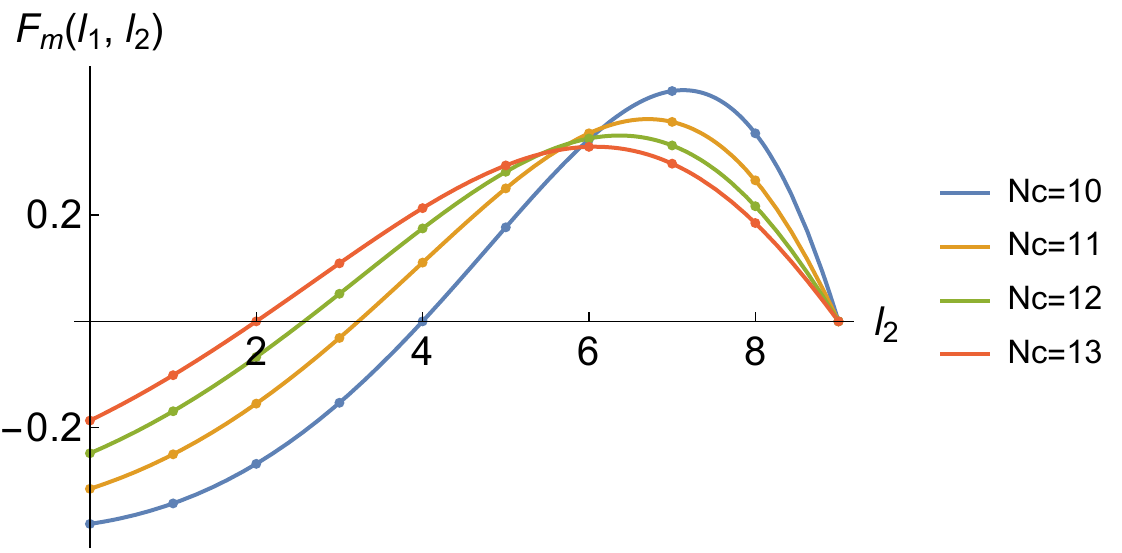}

\vskip 7mm

\includegraphics[scale=0.7]{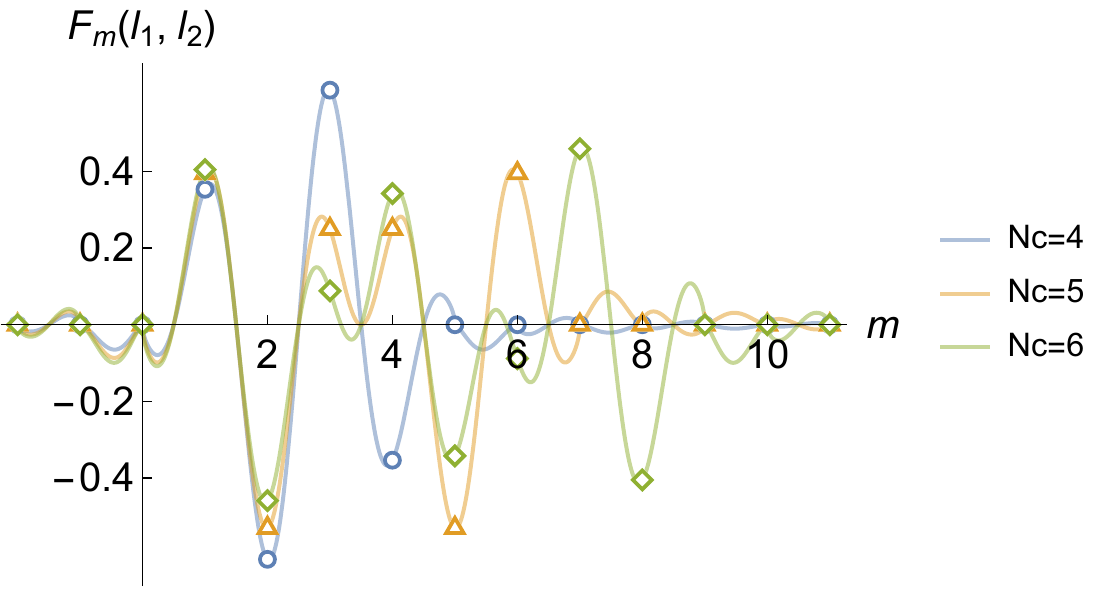}
\hspace{7mm}
\includegraphics[scale=0.7]{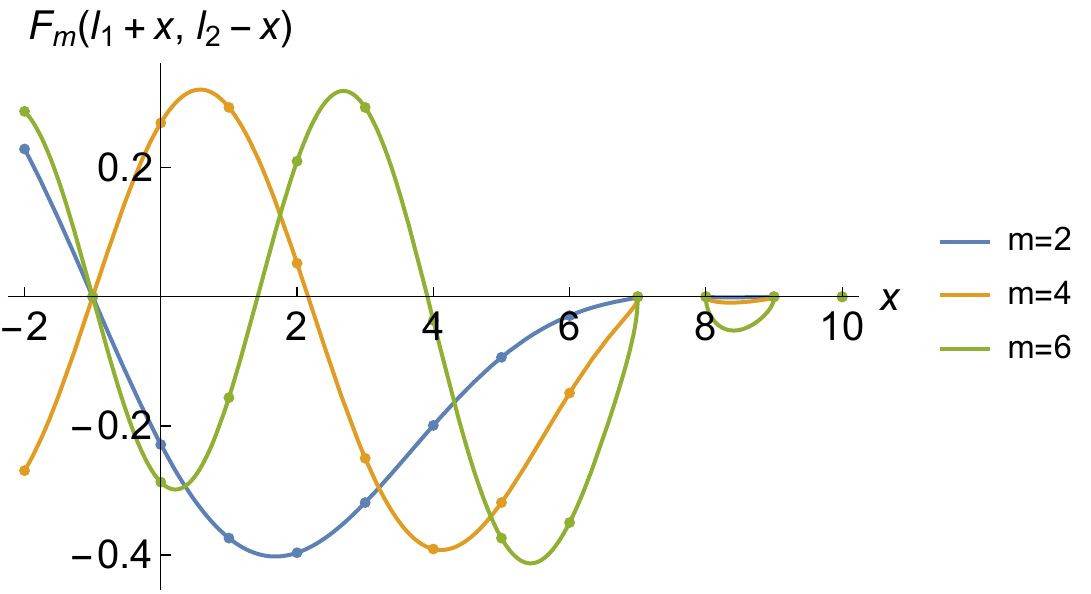}

\caption{Plotting the function $F_m (l_1, l_2)$ against $l_1, l_2$ (above) and $m, x$ (below).}
\label{fig:Fml1l2}
\end{center}
\end{figure}

Generally, the function $F_m (l_1, l_2)$ does not vanish even if $l_2 < 0$, and it slowly decreases to zero when $l_2$ is large and negative. This function has a special zero at\footnote{There are other loci of zeroes, such as $F_{2m} (l_1, 1-N_c)=0$ if $m \le 2 N_c$ and $l_2 \ge N_c -m +2$, which is not meaningful. Practically there is no lower bound for $l_2$\,.}
\begin{equation}
0 = F_{2 m'} (l,l+1) \qquad (l, m' \in \bb{Z}_{\ge 0}) 
\label{Fm special zero}
\end{equation}
meaning that we can solve the Young diagram constraints if all of the following three conditions are satisfied,
\begin{itemize}[nosep]
\item The mode number $m$ is even,
\item $n_Z = l_1 + l_2$ is odd,
\item $n_Z \ge N_c$\,.
\end{itemize}
The last condition may circumvented by using the translation symmetry of the recursion relation\footnote{This is different from the symmetry $\[ \cH_1 \,, \sum_{i=1}^p d_i^\dagger \] = 0$.}
\begin{equation}
\( l_1 \,, l_2 \,, N_c \) \ \to \ \( l_1 + x \,, l_2 +x \,, N_c + x\)
\label{Nc translation}
\end{equation}
although this operation changes the value of $N_c$\,. See Figure \ref{fig:Fml1l2} for the behavior of $F_m(l_1, l_2)$.

One finds that $F_m (l_1,l_2)$ for $l_1, l_2, m \in \bb{Z}_{\ge 0}$ satisfies
\begin{alignat}{9}
0 &= F_m (l_1,l_2) &\qquad &(l_1 > N_c)
\label{F2m boundary-1} \\
0 &= F_m (l_1,l_2) &\qquad &(m<0 \ \ {\rm or} \ \ m > 2 N_c - n_Z + 2)
\label{F2m boundary-2} 
\end{alignat}
as well as \eqref{Fm special zero}. This suggests that the mode number $m$ in the eigenvalue equation \eqref{discrete eqs for Fm} should be chosen from
\begin{equation}
m=2 m', \qquad
m' = 1,2, \dots, \left\lceil N_c - \frac{n_Z}{2} + 1 \right\rceil .
\label{range of m, disc p=2}
\end{equation}
We can derive the identities \eqref{Fm special zero}, \eqref{F2m boundary-1} and \eqref{F2m boundary-2} from the hypergeometric identities in Appendix \ref{app:hypergeom}.

In summary, at $p=2$ we find
\begin{equation}
D_1^G \, O_{F_{2 m'}} = 8 m' \, n_{12} \, \, O_{F_{2m'}} \,, \qquad
O_{F_{2m'}} = \sum_{x=-l_2}^{\lceil (l_1-l_2)/2 \rceil} F_{2m'} (l_1-x, l_2+x) \, O(l_1, l_2) 
\label{app:D1G p=2}
\end{equation}
where $m'$ runs over the range \eqref{range of m, disc p=2}.
The result \eqref{app:D1G p=2} reproduces \cite{deMelloKoch:2012sv}, and is valid for any scaling of $(l_1, l_2)$ with respect to $N_c$\,.
In particular, the eigenvalues are $O(1) \sim O(\lambda/N_c)$ as long as $m \sim O(1)$.

\subsection{Examples of the eigenvalues}

If we introduce a reference point $0$ so that all edges pass through that point, we can write
\begin{equation}
n_{ij} = n_{i0} + n_{j0} \equiv \tilde n_i + \tilde n_j \,, \qquad
\tilde n_E \equiv \sum_{k=1}^p \tilde n_k 
\end{equation}
where $n_E$ is the total number of edges. Then $\scr{N}_{ij}$ simplifies a bit,
\begin{equation}
\scr{N}_{ij} = \tilde n_E \, \delta_{ij} + \tilde{\scr{N}}_{ij} \,, \qquad
\tilde{\scr{N}}_{ij} \equiv
\begin{cases}
(p-2) \, \tilde n_i &\quad (i=j)
\\
- \( \tilde n_i + \tilde n_j \) &\quad (i \neq j) .
\end{cases}
\end{equation}
Consider some special cases. The first case is
\begin{equation}
n_ i = m,
\end{equation}
then
\begin{equation}
\lambda_a = 
\begin{cases}
2 p m &\quad (a=1,2, \dots p-1) \\
0 &\quad (a=p).
\end{cases}
\end{equation}
The residual symmetry is $SO(p-1)$. The second case is
\begin{equation}
\tilde n_1 \gg 1,\quad \tilde n_i \sim O(1) \ \ {\rm for } \ \ (i \neq 1).
\end{equation}
Then
\begin{equation}
\scr{N}_{ij} \sim
\begin{cases}
(p-1) \tilde n_1 &\quad (i=j=1) \\
\tilde n_1 &\quad (j=j \ge 2) \\
- \tilde n_1 &\quad (i=1, j \neq 1 \ {\rm or} \ i \neq 1, j=1) 
\end{cases}
\end{equation}
whose eigenvalues are
\begin{equation}
p \tilde n_1 \,, \ \underbrace{\tilde n_1 , \ \tilde n_1 , \dots , \ \tilde n_1}_{p-2} \,, \ 0
\end{equation}
neglecting $O(1)$ corrections. The residual symmetry is $SO(p-2)$.

\section{Identities of hypergeometric functions}\label{app:hypergeom}

A special case of Gauss hypergeometric function ${}_2 F_1$ is called the Kravchuk (or Krawtchouk) polynomial used in \cite{APW05},
\begin{equation}
K_n (x; p, q) \equiv {}_2F_1 \Big( -n, -x ; -q ; 1/p \Big)
= \frac{1}{\binom{n-q-1}{n}} \, \sum_{j=0}^n p^{-j} \binom{x}{j} \binom{n-q-1}{n-j}
\label{def:Kravchunk}
\end{equation}
which satisfies the recursion relation
\begin{equation}
0 = n (1-p) K_{n-1} (x; p,q)+ \pare{ x - n (1-p) - p (q-n)  } K_n (x ;p,q)
+p (q-n) K_{n+1} (x; p,q) .
\end{equation}

The function ${}_2 F_1 (a,b;c;2)$ is related to ${}_2 F_1 (a,b;c;-1)$ by 
\begin{multline}
{}_2 F_1 (a,b;c;z) = \frac{\Gamma(c) \, \Gamma(c-a-b)}{\Gamma(c-a) \, \Gamma(c-b)} \, {}_2 F_1 (a,b; a+b+1-c ; 1-z)
\\
+ \frac{\Gamma(c) \, \Gamma(a+b-c)}{\Gamma(a) \, \Gamma(b)} \, (1-z)^{c-a-b} {}_2 F_1 (c-a,c-b; 1+c-a-b ; 1-z) .
\end{multline}
As a corollary, 
\begin{equation}
{}_2 F_1 (-a,b;c;z) = \frac{(c-b)_a}{(c)_a} \, {}_2 F_1 (-a, b ; b-c-a+1; 1-z), \qquad (a=0,1,2,\dots)
\end{equation}
where $(x)_a = \Gamma(x+a)/\Gamma(x)$.
We also have
\begin{equation}
{}_2 F_1 (a,b;1+a-b;-1) = \frac{\Gamma(1+a-b) \, \Gamma(1+\frac{a}{2})}{\Gamma(1+a) \, \Gamma (1+\frac{a}{2}-b)}
\end{equation}

\bigskip
Now we can rewrite the rotation matrix \eqref{def:su2 rotation mat} as
\begin{multline}
{}_1 \vev{ j , j_1 \, \big| \, j, j_3} 
= \frac{(-1)^{j+j_3}}{2^j} \, \sqrt{ \binom{2j}{j+j_3} \binom{2j}{j+j_1} } \ \times
\\[1mm]
\frac{(j_1-j)_{j+j_3} }{(-2j)_{j+j_3} } \ 
{}_2 F_1 (-j-j_3, -j-j_1 ; -j_3-j_1+1; -1 ) .
\label{app:su2 rotation mat-2}
\end{multline}
and \eqref{def:Fml} as
\begin{multline}
F_m (l_1,l_2) =
(-1)^{m-1} \, 2^{-N_c + \frac{1}{2} (n_Z-1)} 
\sqrt{\binom{2 N_c-n_Z+1}{m-1} \binom{2 N_c-n_Z+1}{N_c-l_1}} \ \times
\\[1mm]
\frac{(l_2-N_c-1)_{m-1}}{(n_Z-2N_c-1)_{m-1}} \ 
{}_2F_1\left( -m+1, -N_c+l_1 ; N_c - l_2 - m +3 ; -1 \right) .
\label{app:Fml-2}
\end{multline}
Note that the original expression in terms of ${}_2 F_1 (a,b;c;2)$ is more suitable for numerical evaluation.
By combining the above identities, we find
\begin{multline}
F_{2m'} (s,s+1) = 2^{-3 N_c-2+2 m'+3 s} \sqrt{\binom{2 N_c-2 s}{N_c-s} \binom{2 N_c-2 s}{2 m'-1}} \ \times
\\[1mm]
\frac{ \Gamma \(s-N_c+\frac{1}{2}\)  \Gamma \(-N_c-1+2 m'+s\) \Gamma \(N_c+2-2 m'-s\)}{\Gamma (1-m') \Gamma \(-2 N_c-1+2 m'+2 s\) \Gamma \(N_c+\frac{3}{2}-m'-s\)} 
\end{multline}
which vanishes at $m'=0,1,2, \dots$ due to $\Gamma(1-m')=\infty$.

\section{Details of strong coupling analysis}\label{app:detailsStrong}

\subsection{Geometry}\label{app:geometry}

We parametrize S$^5$ by using the embedding coordinates as
\begin{alignat}{9}
X_1 &= R / \sqrt{\rho} \, \cos \eta \, \cos \theta_1
&\qquad\qquad\qquad
X_2 &= R / \sqrt{\rho} \, \cos \eta \, \sin \theta_1
\notag\\[1mm]
X_3 &= R / \sqrt{\rho} \, \sin \eta \, \cos \theta_2
&\qquad\qquad\qquad
X_4 &= R / \sqrt{\rho} \, \sin \eta \, \sin \theta_2
\label{S5 coordinates rho}\\[1mm]
X_5 &= R \sqrt{1 - 1/\rho} \, \cos \phi
&\qquad\qquad\qquad
X_6 &= R \sqrt{1 - 1/\rho} \, \sin \phi 
\notag
\end{alignat}
where $\rho \ge 1, \ \eta \in [0, \pi/2]$ and $\phi, \theta_1 \,, \theta_2 \in [0, 2\pi)$.
The polar coordinates on S$^3$ spanned by $(X^1, X^2, X^3, X^4)$ make the symmetry of $SO(4) = SU(2) \times SU(2)$ manifest.
The metric on $\bb{R}_t \times {\rm S}^5$ is
\begin{equation}
ds^2 = R^2 \pare{ - dt^2 
+ \frac{d \rho^2}{4 (\rho -1) \rho^2}
+ \frac{\(\rho -1 \) d \phi^2}{\rho}
+ \frac{d \eta^2 + \cos^2 \eta \, d \theta_1^2 + \sin^2 \eta \, d \theta_2^2}{\rho} 
}
\label{ds2 rho}
\end{equation}
and the Laplacian is
\begin{equation}
\begin{aligned}
\Delta &= \frac{1}{R^2} \Biggl\{
- \frac{\partial^2}{\partial t^2}
+ 4 (\rho -1) \rho^2 \, \frac{\partial^2}{\partial \rho^2}  
+ 4 \rho \, \frac{\partial}{\partial \rho}  
+\frac{\rho}{\rho -1} \, \frac{\partial^2}{\partial \phi^2} 
+ \frac{\rho }{R^2} \, \Delta_{{\rm S}^3} \Biggr\}
\\[1mm]
\Delta_{{\rm S}^3} &=
\frac{\partial^2}{\partial \eta^2}
+ 2 \cot (2 \eta) \, \frac{\partial}{\partial \eta}
+ \frac{1}{\cos^2 \eta} \, \frac{\partial^2}{\partial \theta_1^2}
+\frac{1}{\sin^2 \eta} \, \frac{\partial^2}{\partial \theta_2^2} \,.
\end{aligned}
\label{app:Laplacian}
\end{equation}
The Jacobian is
\begin{equation}
d \Omega_5 = \frac{R^5 \sin (2 \eta)}{4 \rho^3} \,
d\rho \, d\phi \, d \eta \, d \theta_1 d \theta_2 
\label{Jacobian rho}
\end{equation}
and thus
\begin{equation}
T_3 \int_{\Sigma_4} \frac{R^5 \sin (2 \eta)}{2 \rho^2} \, d\phi \, d \eta \, d \theta_1 d \theta_2 
= N_c \int_{\bb{R}} dt \, \frac{\partial_t \phi}{\rho^2}
\label{C4form rho}
\end{equation}
where \eqref{def:T3} is used.
The RR four-form satisfies $d C_4 \propto (d \Omega_5 + \ast d \Omega_5)$, and the normalization is chosen so that the BPS configuration satisfies $E=J$.

\subsection{Spherical harmonics}\label{app:sph harm}

The scalar spherical harmonics on S$^3$ is \cite{BI66,LachiezeRey:2005hs}
\begin{equation}
\Delta_{{\rm S}^3} \Phi_{k ,m_1,m_2} = - k (k+2) \, \Phi_{k ,m_1,m_2}
\label{app:spherical harmonics S3}
\end{equation}
where $\Delta_{{\rm S}^3}$ is given in \eqref{app:Laplacian} and
\begin{equation}
\begin{gathered}
\Phi_{k ,m_1,m_2} (\eta, \theta_1, \theta_2) = C_{k, m_1, m_2} \, 
\( e^{i \theta_1} \cos \eta \)^{m_1-m_2} \,
\( e^{i \theta_2} \sin \eta \)^{m_1+m_2} \, 
P^{m_1+m_2, m_1-m_2}_{k/2 - m_1} (\cos 2 \eta)
\\[2mm]
C_{k ,m_1,m_2} = \sqrt{ \frac{k+1}{2 \pi^2} } \,
\sqrt{ \frac{(k/2 + m_1)! \, (k/2-m_1)!}{(k/2 + m_2)! \, (k/2-m_2)!} } \,,\qquad
\frac{k}{2} - m_i = 0, 1, \dots, k , \qquad
k \in \bb{Z}_{\ge 0}
\end{gathered}
\label{def:spH3}
\end{equation}
and $P^{a,b}_n (x)$ is the Jacobi polynomial.\footnote{Our convention is same as {\tt JacobiP[n,a,b,x]} in {\tt Mathematica}.}
Note that $\{ m_i \}$ are half-integers when $k$ is odd, and $\Phi_{k ,m_1,m_2}$ is complex-valued,
\begin{equation}
( \Phi_{k ,m_1,m_2} )^* = (-1)^{m_1+m_2} \, \Phi_{k ,-m_1,-m_2} \,.
\end{equation}
The spherical harmonics satisfy the orthogonality
\begin{equation}
\int_0^{2\pi} d \theta_1 \int_0^{2\pi} d \theta_2 \int_0^{\pi/2} d \eta \, \frac{\sin 2 \eta}{2} \,
(-1)^{m_1+m_2} \, \Phi_{k ,m_1,m_2} \, \Phi_{l ,-n_1, -n_2}
= \delta_{l, k} \, \delta_{m_1,n_1} \, \delta_{m_2,n_2} 
\label{ON Sph Harm}
\end{equation}
and the integrated spherical harmonics satisfy
\begin{multline}
\int_0^{2\pi} d \theta_1 \int_0^{2\pi} d \theta_2 \int_0^{\pi/2} d \eta \frac{\sin 2 \eta}{2} \,
\Bigl\{
(\partial_\eta \Phi_{k ,m_1,m_2}) \, (\partial_\eta \Phi_{l ,-n_1, -n_2})
+ \frac{(\partial_{\theta_1} \Phi_{k ,m_1,m_2}) \, (\partial_{\theta_1} \Phi_{l ,-n_1, -n_2})}{\cos^2 \eta}
\\[1mm]
+ \frac{(\partial_{\theta_2} \Phi_{k ,m_1,m_2}) \, (\partial_{\theta_2} \Phi_{l ,-n_1, -n_2})}{\sin^2 \eta}
\Bigr\}
= k (k+2) \delta_{l, k} \, \delta_{m_1,n_1} \, \delta_{m_2,n_2} \,.
\label{identity int SpH}
\end{multline}

The first few eigenfunctions are
\begin{gather}
\Phi_{0,0,0} = \frac{1}{\sqrt{ 2 \pi^2} } \,, \qquad
\Phi_{1,\frac12,\frac12} = \frac{e^{i \theta_2} \sin \eta}{\pi} \,, \qquad
\Phi_{1,\frac12,-\frac12} = \frac{e^{i \theta_1} \cos \eta}{\pi} \,,
\\[1mm]
\Phi_{2,0,0} = \sqrt{\frac{3}{2 \pi^2}} \, \cos (2\eta) , \qquad
\Phi_{2,1,0} = \sqrt{\frac{3}{4 \pi^2}} \, e^{i (\theta_1 +\theta_2)} \, \sin (2\eta) , \qquad
\Phi_{2,0,1} = \sqrt{\frac{3}{4 \pi^2}} \, e^{i (\theta_1 -\theta_2)} \, \sin (2\eta) .
\notag
\end{gather}
The functions $\Phi_{2 \ell,0,0}$ are equal to the Legendre polynomial $P_\ell (\cos 2\eta)$.

\subsection{Classical solutions at $j=0,1$}\label{app:special sol}

Consider the classical D3-brane solutions around $j=0,1$, or equivalently $J=0, N_c/g_s$\,.
For this purpose, we use the coordinate $r = 1/\sqrt{\rho}$ with $0 \le r \le 1$.
The metric \eqref{ds2 rho} becomes
\begin{equation}
ds^2 = R^2 \pare{ - dt^2 
+ \frac{dr^2}{1-r^2}
+ (1-r^2) \, d\phi^2 
+ r^2 \Bigl( d \eta^2 + \cos^2 \eta \, d \theta_1^2 + \sin^2 \eta \, d \theta_2^2 \Bigr) 
} .
\label{ds2 r}
\end{equation}
We expand the equations of motion around
\begin{equation}
r = \sqrt{j} + \epsilon \, r_1 (t, \eta), \qquad
\phi = t + \epsilon \, \phi_1 (t, \eta) , \qquad
(j=0,1).
\label{rphi epsilon ansatz}
\end{equation}

\bigskip
First, consider the case $j=1$.
The EoM for $\phi$ becomes $\partial_t r_1 = 0$. 
If we write $r_1 \equiv - e^{-s(\eta)}$, the EoM for $r$ becomes
\begin{multline}
0 = \frac{1}{4} \left(s'{}^2 - 4 \cot (2 \eta ) s' - 2 s''\right)
\\
-\epsilon \pare{ 2 \, \partial_t \phi_1
- \frac{e^{-s(\eta )}}{16}  \left(
6 s'' (s'^2+2) -3 s'^2 (s'^2+4) +4 \cot (2 \eta ) \, s' (s'^2+6) 
\right)
}
+ O(\epsilon^2) .
\label{eom-r j1}
\end{multline}
The general solution at $O(\epsilon^0)$ is
\begin{equation}
e^{-s(\eta)} = \( \frac{c_{(2)} + c_{(1)} \log \cot \eta}{4} \)^2 .
\end{equation}
To maintain $r \in [0,1]$ we need $c_{(1)}=0$. Then, the equation \eqref{eom-r j1} produces $\partial_t \phi_1=0$. Thus there is no non-trivial solution.

\bigskip
Next, consider the case $j=0$. The EoM's give us
\begin{equation}
0 = r_1 \( \cot (2 \eta) r_1 + \partial_\eta r_1 \)
= r_1 \, \partial_\eta \phi_1 \,.
\end{equation}
The non-trivial solution of the first equation is
\begin{equation}
r_1 = \frac{c}{\sqrt{ \sin 2 \eta }}
\end{equation}
for a constant $c$. This function diverges around $\eta=0, \pi/2$, and is inconsistent with $r \in [0,1]$.
Again, we find no non-trivial solution.

\section{On non-abelian DBI}\label{app:non-abelian}

The symmetry of the effective $U(p)$ theory becomes non-abelian if some of the column lengths become equal, $l_i = l_{i+1}$ for some $i$ in \eqref{def:r1R}.
Correspondingly, some of the $p$ giant graviton branes coincide at strong coupling.
Let us make a short digression about non-abelian DBI action to examine this situation.\footnote{The author thanks A. Tseytlin for his comments on abelian and non-abelian DBI.}

The DBI action is a low energy effective action of closed and open string massless modes on the brane.
As a worldvolume theory, the DBI action without $U(1)$ flux is also an example of 4d conformal theory.
In \AdSxS, the conformal symmetry of the target spacetime is nonlinearly realized \cite{Jevicki:1998qs,Kuzenko:2002zh}.
The DBI action can be made supersymmetric in the sense of $\kappa$ symmetry \cite{Aganagic:1996pe,Aganagic:1996nn,Metsaev:1998hf} and of the worldvolume symmetry \cite{Rocek:1997hi,Tseytlin:1999dj}.
The fundamental strings can be coupled to multiple coincident D-branes by introducing non-abelian flux $F_{\mu\nu}$ \cite{Tseytlin:1997csa}. The addition of the CS term to the non-abelian DBI induces dielectric effects \cite{Myers:1999ps}.

One way to define non-abelian DBI is to expand the DBI action in a formal series of $F$,
\begin{equation}
\det (G + 2 \pi \alpha' F) = \det G \( 
1 - \frac{(2 \pi \alpha')^2}{2} \, G^{ab} F_{bc} G^{cd} F_{da} + \dots \) .
\label{expand detG}
\end{equation}
Then we promote $F$ to a non-abelian field, and take the trace. This procedure suffers from the ordering ambiguity, which should be fixed by the consistency with the open string amplitude \cite{Medina:2002nk,Barreiro:2005hv,Oprisa:2005wu,Broedel:2013tta}.

In Section \ref{sec:strong coupling}, we want to find classical solutions continuously connected to the spherical giant graviton. From the above prescription for the non-abelian DBI in \eqref{expand detG}, we do not see any significant difference between $U(1)^p$ and $U(p)$ theories at the order of $\alpha'{}^2$.

The induced metric $G_{ab}$ should not have the $U(p)$ structure for the following reasons.
First, the induced metric comes from closed string massless modes, which do not see non-abelian symmetry.
Second, the attempts to make $G_{ab}$ a matrix-valued field suffer from various difficulties \cite{Douglas:1997ch,DeBoer:2001uk}.
Third, in our deformation problem, we are only interested in the corrections of $O(\alpha'{}^2)$, and the commutator terms do not show up at this order.
Then, all matrices are simultaneously diagonalizable, unless they couple to other objects.

\bibliographystyle{utphys}
\bibliography{bibgg}{}

\end{document}